\documentclass[aps,prl,reprint,superscriptaddress]{revtex4-2}

\usepackage{bm}
\usepackage{graphicx}
\usepackage[caption=false]{subfig}
\usepackage[colorlinks=true, citecolor=blue, linkcolor=blue, urlcolor=blue]{hyperref}
\usepackage{pinlabel} 
\usepackage{amsmath}
\usepackage{dcolumn}

\renewcommand\thesubfigure{(\alph{subfigure})} 

\begin{document}


\title{Suppression of Unwanted $ZZ$ Interactions in a Hybrid Two-Qubit System}



\author{Jaseung Ku}
\affiliation{Department of Physics, Syracuse University, Syracuse, New York 13244, USA}

\author{Xuexin Xu}
\affiliation{Peter Gr\"unberg Institute, Forschungszentrum J\"ulich, J\"ulich 52428, Germany}
\affiliation{J\"ulich-Aachen Research Alliance (JARA), Fundamentals of Future Information Technologies, J\"ulich 52428, Germany}

\author{Markus Brink}
\author{David C. McKay}
\author{Jared B. Hertzberg}
\affiliation{IBM, T. J. Watson Research Center, Yorktown Heights, New York 10598, USA}

\author{Mohammad H. Ansari}
\affiliation{Peter Gr\"unberg Institute, Forschungszentrum J\"ulich, J\"ulich 52428, Germany}
\affiliation{J\"ulich-Aachen Research Alliance (JARA), Fundamentals of Future Information Technologies, J\"ulich 52428, Germany}

\author{B.L.T. Plourde}
\affiliation{Department of Physics, Syracuse University, Syracuse, New York 13244, USA}


\date{\today}

\begin{abstract}
Mitigating crosstalk errors, whether classical or quantum mechanical, is critically important for achieving high-fidelity entangling gates in multi-qubit circuits. For weakly anharmonic superconducting qubits, unwanted $ZZ$ interactions can be suppressed by combining qubits with opposite anharmonicity. We present experimental measurements and theoretical modeling of two-qubit gate error for gates based on the cross resonance interaction between a capacitively shunted flux qubit and a transmon, and demonstrate the elimination of the $ZZ$ interaction.
\end{abstract}


\maketitle

Superconducting qubits are a promising candidate for building fault-tolerant quantum computers~\cite{clarke_superconducting_2008,devoret_superconducting_2013,gambetta_building_2017,krantz_quantum_2019}. However, the gate errors in current devices are not definitively below the threshold required for fault-tolerance. Despite tremendous improvements in qubit coherence, circuit design, and control, two-qubit gate errors remain in the range of $4-9\times10^{-3}$~\cite{arute_quantum_2019, Sheldon_AC_2016}. This is worse than what would be naively expected based on current device coherences~\cite{wei2019verifying}. One limiting factor to these errors is crosstalk in the device corresponding to unwanted terms in the Hamiltonian. This is a particular concern for one of the more common superconducting qubit architectures – fixed-frequency transmons~\cite{koch2007charge} coupled to nearest neighbors via a static exchange term $J$. In this architecture, the two-qubit gate is enabled by activating the cross-resonance (CR) effect~\cite{chow_simple_2011, Corcoles_RB_2013, tripathi_operation_2019}, where a $ZX$ interaction term is generated by driving one qubit (the control) at the frequency of the neighboring qubit (the target).

CR has several advantages: it allows for all-microwave control of a fixed-frequency device, and is thus simple from a control perspective; also, the use of non-tunable qubits removes a source of decoherence. The strength of the CR effect is proportional to $J$~\cite{magesan2018effective}. However, for transmons, which have a negative value of the anharmonicity -- the difference between the primary qubit transition out of the qubit subspace and the qubit transition -- this $J$ also produces an always-on $ZZ$ coupling term.  Such a $ZZ$ interaction, whether static or driven during the CR gate~\cite{magesan2018effective}, is an ever-present source of error. Unlike classical crosstalk, which can be cancelled by the appropriate application of compensation tones~\cite{Sheldon_AC_2016,magesan2018effective}, the $ZZ$ term leads to unwanted entanglement between pairs and so is not easily mitigated unless, for example, additional circuitry, such as a tunable coupler, is added~\cite{Mundata_crosstalk_PRAppl_2019}.  

As an alternative approach, if the transmon qubit can be combined with a qubit design where the anharmonicity is positive, the $ZZ$ term can be cancelled at specific qubit-qubit detunings, and the CR effect between the two qubits utilized to form a high-fidelity gate. Fortunately, such a qubit exists -- the capacitively shunted flux qubit (CSFQ)~\cite{steffen_high-coherence_2010}. Recently, the CSFQ has regained attention, in part, due to its greatly improved coherence time~\cite{yan_flux_2016}. Although the CSFQ is a flux-tunable device, it can be operated at a flux sweet spot (flux bias $f=\Phi/\Phi_0=0.5$, where $\Phi_0=h/2e$, $h$ is Planck's constant, and $e$ is the electron charge), where it is first-order insensitive to flux noise. The anharmonicity at the sweet spot can be positive and large ($>$~+500~MHz), which provides a parameter regime that is otherwise inaccessible in all-transmon devices~\cite{yan_tunable_2018,zhao2020highcontrast}. 

In this manuscript we present measurements of the first such hybrid CSFQ-transmon device and theoretical modeling to investigate its performance. First, we experimentally demonstrate and theoretically model the suppression of the static $ZZ$ interaction for a particular detuning of the CSFQ and transmon. Second, we investigate the characteristic behavior of the CR effect as a function of CSFQ-transmon detuning. Third, we explore the dependence of two-qubit gate error on both flux and gate length. Finally, we use our model to describe the requirements for a future device capable of achieving a two-qubit gate error of $1\times 10^{-3}$. 
\begin{figure}
\begin{subfloat}{\label{fig:circuit_schematics}}
\labellist
\bfseries
\pinlabel (a) at -2 200
\endlabellist
\includegraphics[width=0.95\columnwidth]{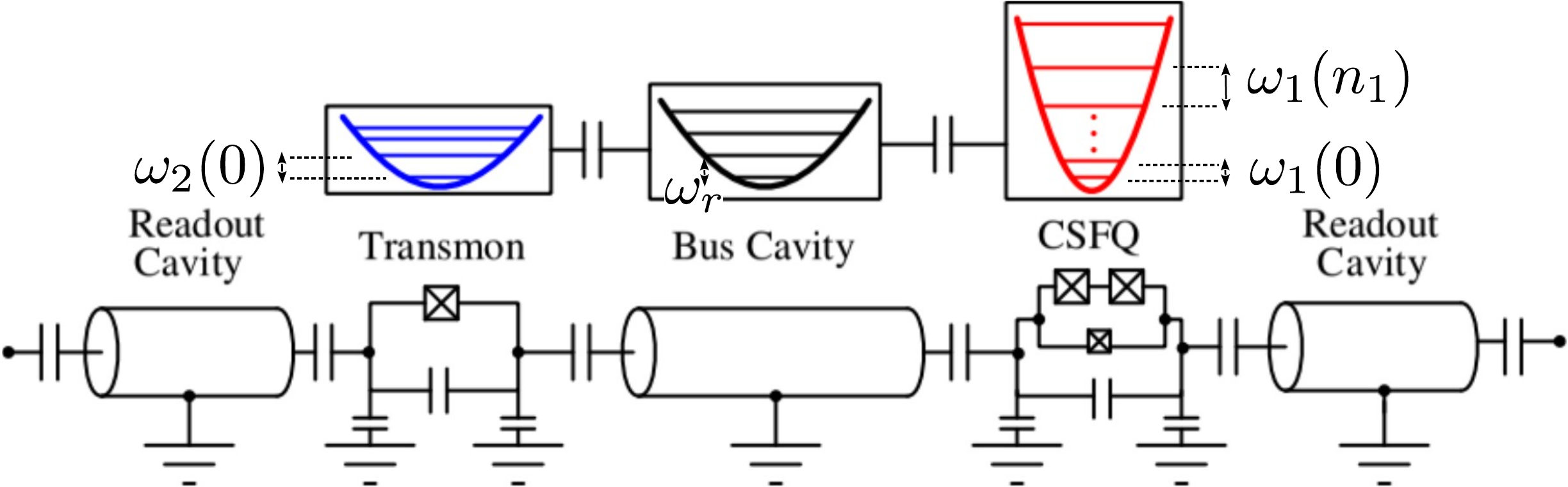}
\end{subfloat}
\\[0.3cm]
\begin{subfloat}{\label{fig:freq_diagram}}
\labellist
\bfseries
\pinlabel (b) at 15 450
\pinlabel (c) at 350 450
\endlabellist
\includegraphics[width=\columnwidth]{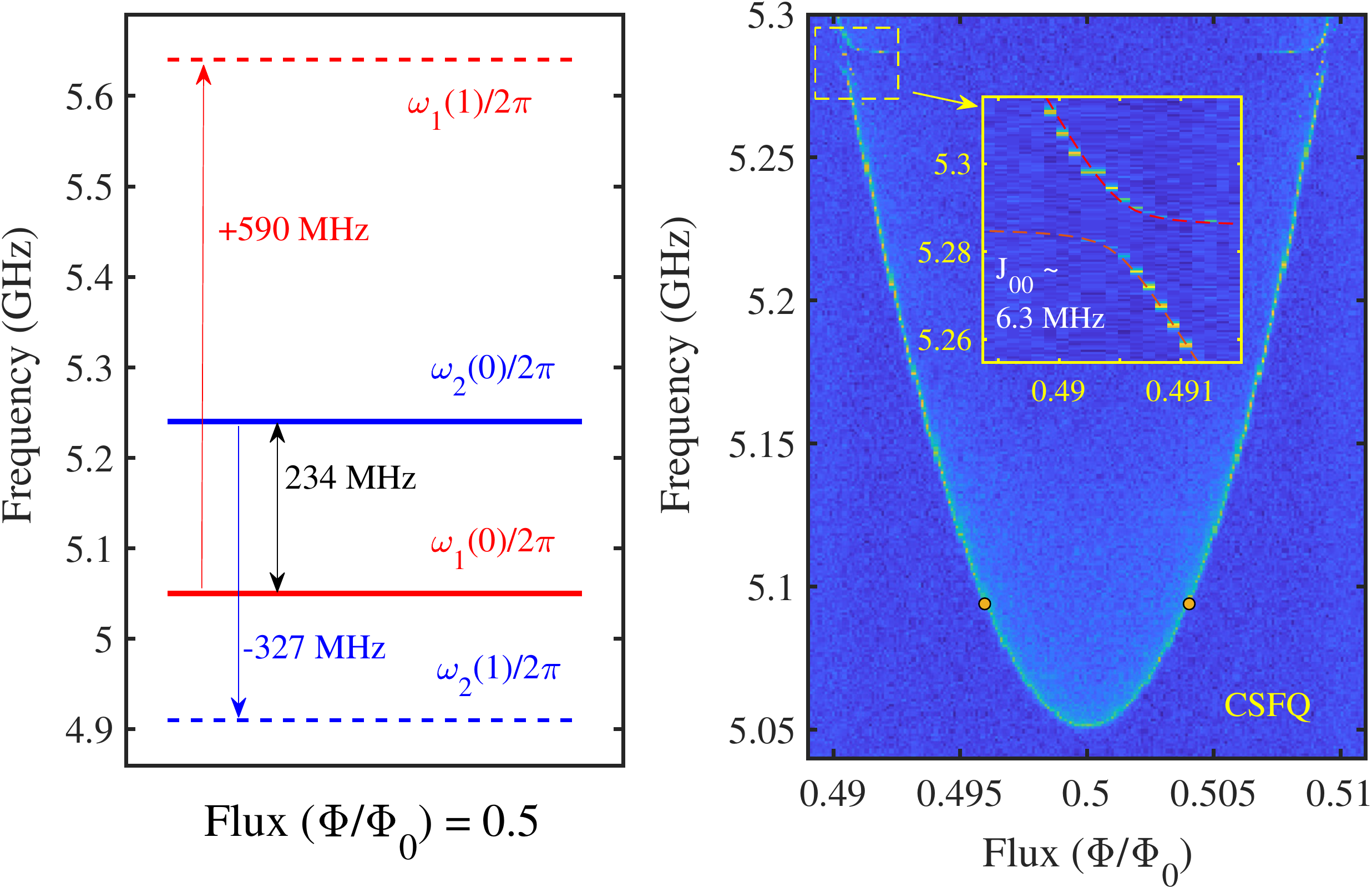}
\end{subfloat}
\begin{subfloat}{\label{fig:CSFQ_spec}}
\end{subfloat}
\vspace{-0.24in}
\caption{\label{fig:schematic} (a) Simplified circuit diagram of CSFQ-transmon system coupled via a bus cavity. Potential energies with eigenenergies for transmon, bus cavity, and CSFQ are depicted above the circuit. (b) Frequency diagram of transmon and CSFQ at flux sweet spot. (c) CSFQ qubit frequency spectrum vs. external magnetic flux. Orange dots at $f=0.496, 0.504$ indicate the flux points where the static $ZZ$ becomes zero. (Inset) Anti-crossing of transmon and CSFQ with fit (red dashed line).
}
\end{figure}

The device consists of a fixed-frequency transmon and CSFQ coupled via a bus cavity resonator [Fig.~\hyperref[fig:schematic]{\ref{fig:circuit_schematics}}]. Each qubit has its own readout resonator with a microwave input/output port. Details on sample fabrication, measurement setup, and device parameters can be found in the Supplement~\cite{suppl}. This coupled two-qubit system can be described by the Hamiltonian:
\begin{equation}\label{eq:two-qubit_Hamil}
\begin{split}
    H&=\sum_{q=1,2}\sum_{n_q}{\omega}_q(n_q)\left|n_q\right>\left<n_q\right|+\sqrt{(n_1+1)(n_2+1)}\\
    &\times J_{n_1,n_2}\left(\left|n_1+1,n_2\right>\left<n_1,n_2+1\right|+h.c.\right),
\end{split}
\end{equation}
where $\omega_q (n_q)$ is the bare transition frequency between energy levels $n_q$ and $n_q+1$ for qubit $q$. The primary qubit transition is thus $\omega_q(0)$ and we define $\omega_q\equiv\omega_q(0)$. The coupling strength  $J_{n_1,n_2}$  provides an indirect two-photon interaction via a bus resonator between energy levels $n_1$ and $n_1+1$ in qubit 1 and levels $n_2$ and $n_2+1$ in qubit 2 (see Supplement~\cite{suppl} for details). We take $\hbar=1$ throughout.

The qubits were measured using conventional circuit-QED techniques in the dispersive regime~\cite{blais_cavity_2004}. The measured qubit frequency, anharmonicity, and qubit-qubit detuning for the CSFQ and transmon at the sweet spot are shown in Fig.~\hyperref[fig:schematic]{\ref{fig:freq_diagram}}. The tunability of the CSFQ spectrum as a function of flux [Fig.~\hyperref[fig:schematic]{\ref{fig:CSFQ_spec}}] allows us to explore a range of qubit-qubit detuning in the following experiments. We fit the anticrossing between the CSFQ and transmon [Fig.~\hyperref[fig:schematic]{\ref{fig:CSFQ_spec}} inset] to obtain the zeroth-order exchange coupling strength $J_{00}/2\pi= 6.3$~MHz.  The average single-qubit gate fidelity was measured with the standard randomized benchmarking (RB) protocol (details in Supplement~\cite{suppl}), giving the average gate error lower than $1\times10^{-3}$. For a CR drive, we take the CSFQ (transmon) as the control (target) qubit. 

\begin{figure}
\centering
\includegraphics[width=0.95\columnwidth]{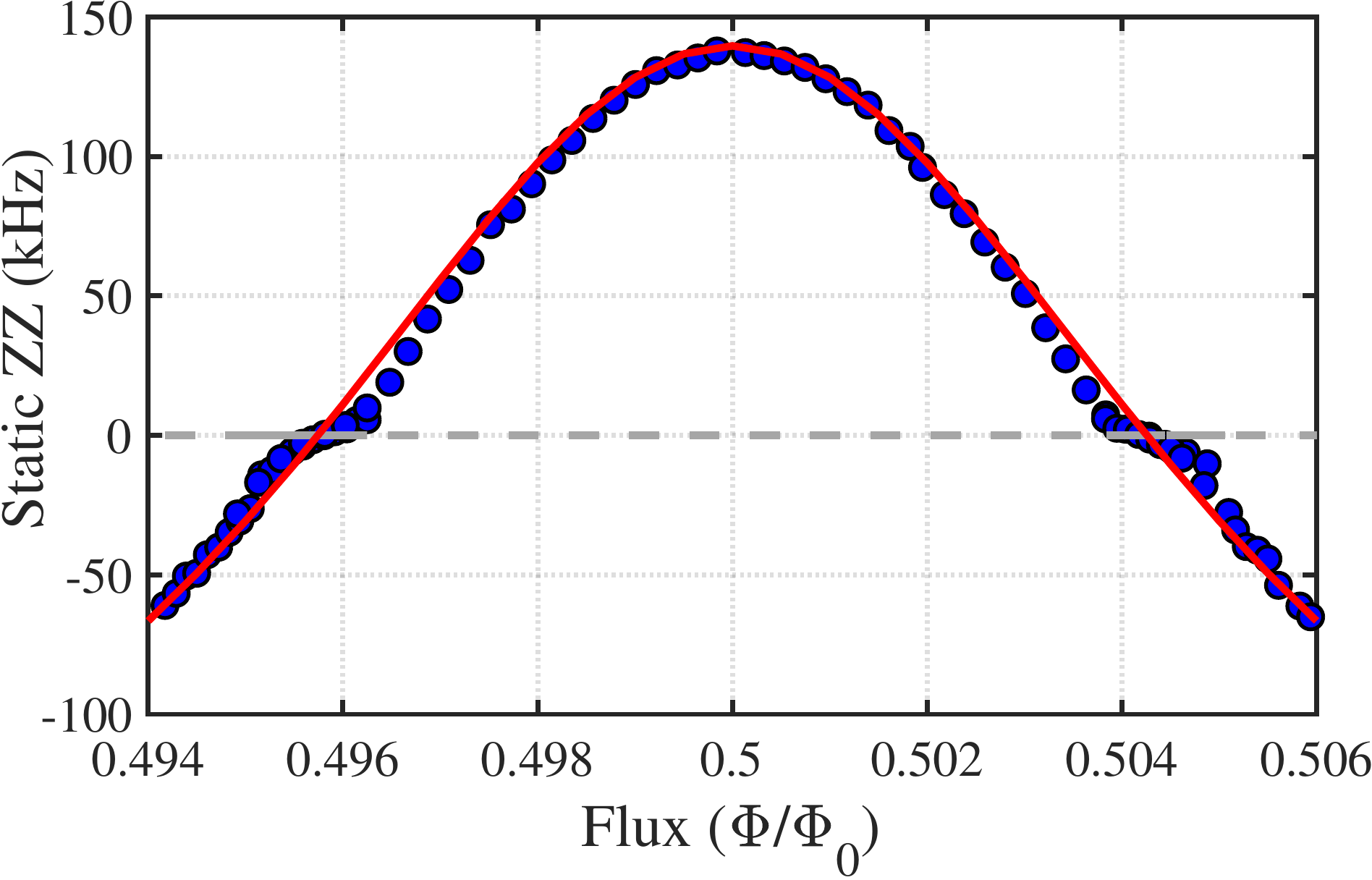}
\vspace{-0.14in}
\caption{\label{fig:ZZ_flux} Static $ZZ$ measured as a function of flux via Joint Amplification of $ZZ$ (JAZZ) protocol~\cite{takita_experimental_2017}. The static $ZZ$ becomes zero at two flux points $\Phi/\Phi_0=0.496, 0.504$. The size of the error bars is comparable with or smaller than the size of the data symbols. The red solid line represents a theory calculation using Eq.~\eqref{eqn:ZZ}.}
\end{figure}
We investigate how the static $ZZ$ interaction of the system varies with the flux bias of the CSFQ. The effective Hamiltonian that is diagonal in the dressed frame is,
\begin{equation}\label{eqn:twoQ_eff_Hamil}
    H_{\rm{eff}}=-\tilde{\omega}_1\frac{ZI}{2}-\tilde{\omega}_2\frac{IZ}{2}+\zeta\frac{ZZ}{4},
\end{equation}
where $\tilde{\omega}_1$ and $\tilde{\omega}_2$ are the dressed qubit frequencies. $\zeta$ is the frequency shift of one qubit when the other qubit is excited from the ground state: $\zeta=(E_{11}-E_{10})-(E_{01}-E_{00})$, where $E_{ij}$ is the energy eigenvalue of the Hamiltonian for qubit 1 at $|i\rangle$ and qubit 2 at $|j\rangle$. The static $ZZ$ interaction arises when higher energy levels are involved in the two-qubit Hamiltonian. $ZZ$ interaction results in an additional phase rotation depending on the state of either qubit, thus contributing to two-qubit gate error. For our device, the static $ZZ$ strength has a maximum value of 140~kHz at the flux sweet spot, but away from this point it decreases and eventually crosses zero near $\Phi/\Phi_0$ = 0.496 and 0.504 (Fig.~\hyperref[fig:ZZ_flux]{\ref{fig:ZZ_flux}}), where the CSFQ-transmon detuning is 191~MHz. $ZZ$-free qubit pairs can be obtained if $\zeta$ vanishes in Eq.~\eqref{eqn:twoQ_eff_Hamil}. A detailed analysis involving block-diagonalization of the multilevel Hamiltonian [Eq.~\eqref{eq:two-qubit_Hamil}] into the qubit subspace shows that $\zeta$ can be expressed as (see Supplement~\cite{suppl} for details):
\begin{eqnarray}\label{eqn:ZZ}
\zeta &=& -\frac{2J^2_{01}}{\Delta + \delta_{2}} + \frac{2J^2_{10}}{\Delta-\delta_{1}},
\end{eqnarray}
where $\Delta=\omega_2-\omega_1$ is the qubit-qubit detuning, and $\delta_i=\omega_i(1)-\omega_i$ is the anharmonicity of qubit $i$. Within the limit $|\Delta|<|\delta|$, where the CR effect is strongest~\cite{ware2019crossresonance}, for a transmon-transmon device, both terms of Eq.~\eqref{eqn:ZZ} are positive, and thus $ZZ$ interactions will always be present in all-transmon circuits with fixed couplings. However, in a CSFQ-transmon circuit the second term in Eq.~\eqref{eqn:ZZ} can be negative, due to the large and positive anharmonicity of the CSFQ. This allows the hybrid CSFQ-transmon system to be static $ZZ$-free. Eq.~\eqref{eqn:ZZ} was used to compute the flux dependence of the static $ZZ$ strength using separately extracted device parameters, including the flux-dependent anharmonicity and transition frequencies of the CSFQ (red solid line in Fig.~\hyperref[fig:ZZ_flux]{\ref{fig:ZZ_flux}}). The agreement between theory and experiment is quite good except near the zero-crossing points, where the experimental $ZZ$ data exhibits a kink. We speculate that this could be due to the breakdown of our perturbative treatment of the effective Hamiltonian, and thus Eq.~\eqref{eqn:ZZ}. Away from the flux sweet spot, the qubit-qubit detuning decreases, while $J_{10}$ increases, thus pushing the ratio $J/\Delta$ beyond the dispersive limit. A framework for treating such situations is discussed in Ref.~\onlinecite{ansari_superconducting_2019}.

\begin{figure}
\centering
\includegraphics[width=0.9\columnwidth]{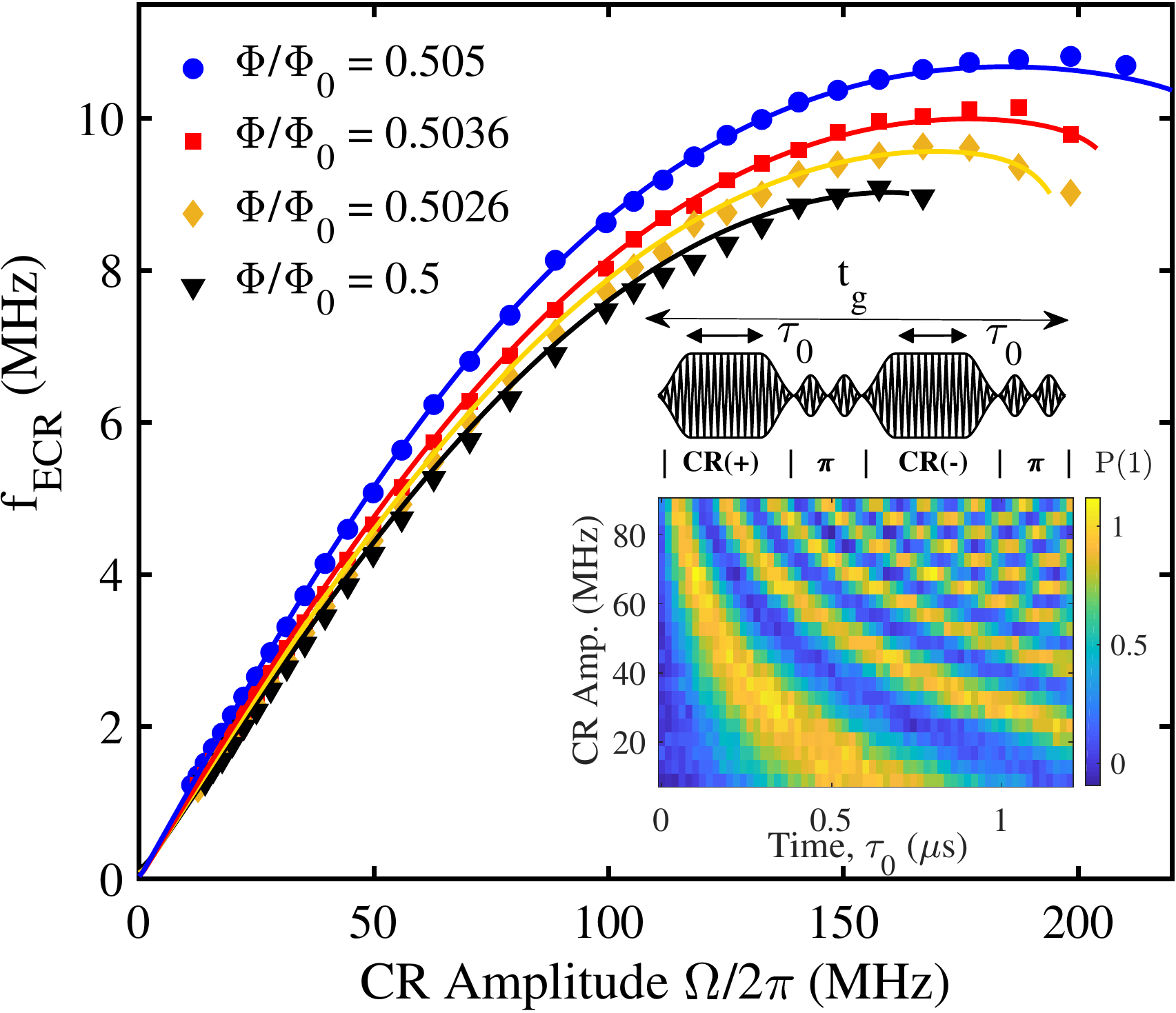}
\vspace{-0.05in}
\caption{ Echoed CR rate vs. CR amplitude at four representative flux points. The corresponding qubit-qubit detunings are (234, 217, 199, 166) in MHz. Solid lines correspond to theoretical model. (Inset) Color density plot of the oscillation of target qubit driven with various CR amplitudes at the flux sweet spot. Colorbar represents the first excited state probability of the target qubit. Echoed CR pulse sequence is illustrated above the density plot.}
\label{fig:CR_flux}
\end{figure}
For the CR effect, a drive tone applied to the control qubit at the frequency of the target qubit induces a rotation of the target qubit with the direction of rotation dependent on the state of the control qubit, thus corresponding to a $ZX$ term in the effective Hamiltonian~\cite{magesan2018effective}. Due to terms other than $ZX$ in the full CR Hamiltonian, an echoed CR protocol is commonly used, which removes $ZI$ and $IX$ contributions~\cite{Corcoles_RB_2013}. We performed echoed CR to measure the rotation rate, $f_{\rm ECR}$, as a function of CR amplitude at different flux points (Fig.~\hyperref[fig:CR_flux]{\ref{fig:CR_flux}}). The echoed CR pulse consists of two Gaussian flat-top CR pulses with $\pi$ phase difference, and a $\pi$-pulse on the control qubit after each CR pulse (Fig.~\ref{fig:CR_flux} inset). We define the two-qubit gate length $t_g=2\tau_0+160$~ns, where $\tau_0$ is the flat-top length of each CR pulse; the constant 160~ns corresponds to the sum of the rising/falling edges on the CR pulses and the $\pi$ pulses applied to the control qubit. With variable $\tau_0$, the oscillation frequency of the transmon was measured for a range of CR amplitude (Fig.~\hyperref[fig:CR_flux]{\ref{fig:CR_flux}} inset). The CR amplitude was calibrated in terms of the Rabi frequency of the CSFQ at the flux sweet spot. The echoed CR rate increases almost linearly at low CR amplitude, while for the stronger CR drive it slows down as the CSFQ is driven off-resonance~\cite{chow_simple_2011}. Eventually, the rate levels off to a maximum as the energy levels $E_{11}$ and $E_{02}$ get closer and finally anticross each other at the CR amplitude corresponding to the maximum. Applying a non-perturbative diagonalization scheme to the effective Hamiltonian [Eq.~\eqref{eqn:twoQ_eff_Hamil}] together with a CR driving Hamiltonian, we simulated $f_{\rm ECR}$ vs. CR amplitude (details in Supplement~\cite{suppl}). The resulting theoretical curves for $f_{\rm ECR}$ vs. CR amplitude agree well with the experimental points (Fig.~\hyperref[fig:CR_flux]{\ref{fig:CR_flux}}).

\begin{figure}
\centering
\includegraphics[width=0.95\columnwidth]{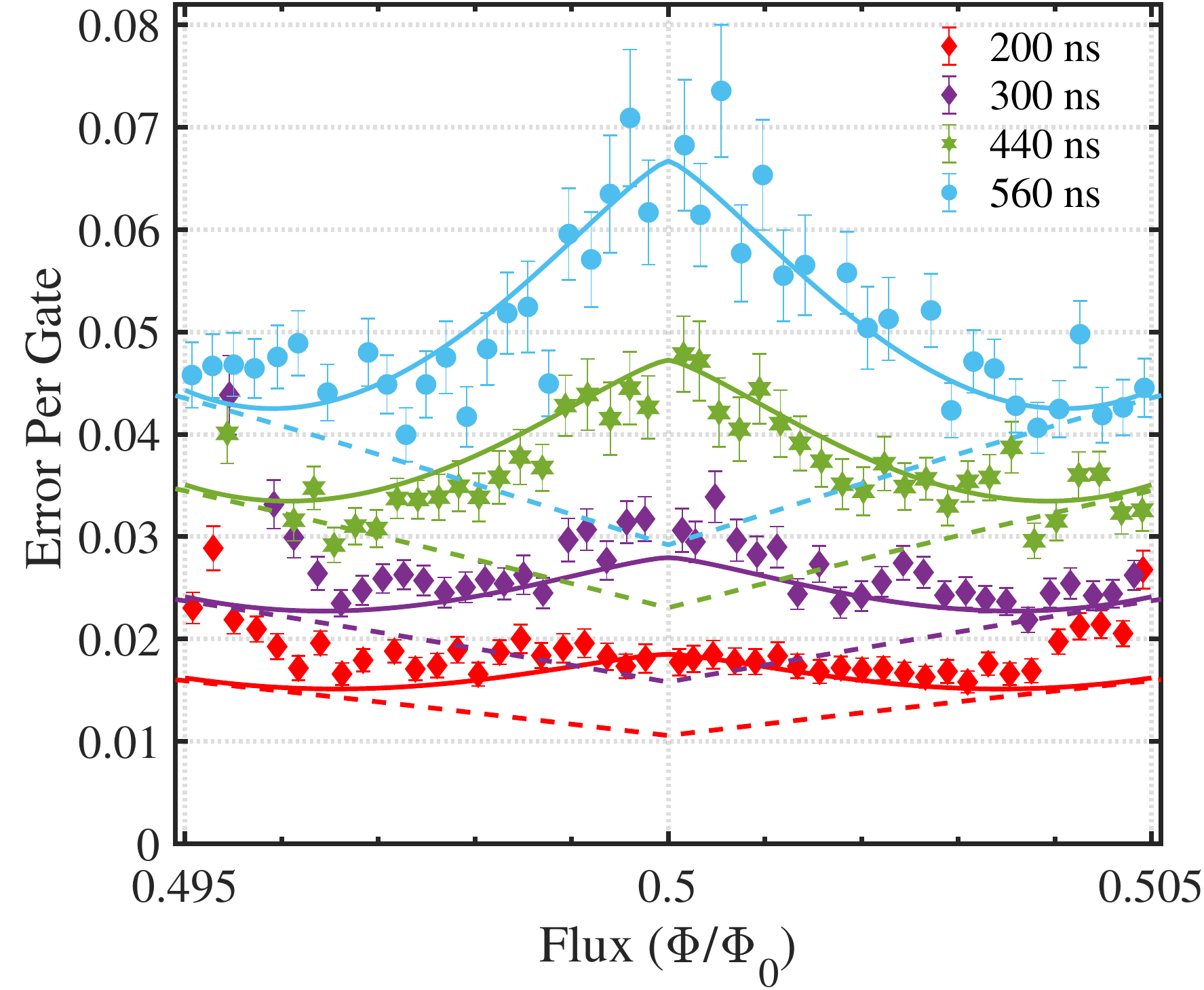}
\vspace{-0.1in}
\caption{\label{fig:twoQ_fid_flux} Average error per two-qubit gate plotted versus flux for four representative two-qubit gate lengths. Dashed lines indicate theoretical coherence-limited two-qubit gate errors with no $ZZ$ interactions; full theory simulations are shown by solid lines.
}
\end{figure}

The average two-qubit error per gate was measured via standard randomized benchmarking (RB)~\cite{Magesan_RB_PRL2011} at various flux points and gate lengths $t_g$ of the  $ZX_{90}$, which serves as the pulse primitive for the two-qubit entangling gate~\cite{Corcoles_RB_2013} (Fig.~\hyperref[fig:twoQ_fid_flux]{\ref{fig:twoQ_fid_flux}}). For each flux point, the primitive single-qubit gate ($X_{90}$) and two-qubit gate ($ZX_{90}$) were re-calibrated. No active cancellation pulse for removing classical crosstalk~\cite{Sheldon_AC_2016} was used. The RB data was fit to the standard fidelity decay curve $A\alpha^m +B$, where $m$ is the number of Clifford gates and $\alpha$ the depolarization parameter~\cite{Magesan_RB_PRL2011}. The average two-qubit error per gate $\epsilon$ was then  calculated using the expression, $\epsilon=(3/4)\cdot(1-\alpha^{1/N})$, where $N$ is the average number of $ZX_{90}$ gates per two-qubit Clifford gate~\cite{mckay_efficient_2017, mckay_three-qubit_2019}. 

By increasing the gate length, a characteristic ``W"-shaped pattern develops with respect to flux, corresponding to larger errors at the sweet spot with minima to either side, followed by increasing error for further flux biasing away from 0.5. The smallest gate error, $1.6\times10^{-2}$, occurs for $t_g = 200$~ns and $f = 0.496, 0.504$ (Fig.~\hyperref[fig:twoQ_fid_flux]{\ref{fig:twoQ_fid_flux}}). This behavior can be described by the interplay between fidelity loss from the $ZZ$ interaction and classical crosstalk on the one hand, and fidelity gain from longer coherence times near the sweet spot on the other hand. Away from the sweet spot, the $ZZ$ interaction and classical crosstalk decrease and the gate fidelity approaches the coherence limit. Including the $ZZ$ interaction and classical crosstalk in our simulation was sufficient to reproduce the flux-dependence of the experimental gate errors.

The dashed lines in Fig.~\hyperref[fig:twoQ_fid_flux]{\ref{fig:twoQ_fid_flux}} correspond to the coherence-limited gate error, which is mainly dominated by the CSFQ's $T_2$. Due to flux noise, the CSFQ has a maximum $T_2$ at the sweet spot, which quickly decreases away from this point (see $T_2$ vs. flux in Supplement~\cite{suppl}). As is clear from Fig.~\hyperref[fig:twoQ_fid_flux]{\ref{fig:twoQ_fid_flux}}, the coherence-limit curves alone are not sufficient to reproduce the measured flux-dependence of the gate error. The static $ZZ$ strength (Fig.~\ref{fig:ZZ_flux}) has a significant impact on the gate error, and was included in the simulation. Moreover, we model classical crosstalk in a similar manner to Ref.~\onlinecite{magesan2018effective}, by including in the CR driving Hamiltonian a modified amplitude $R(f,t_g)\Omega$ and shifted phase, where $R(f, t_g)$ is a scaling factor. $R$ was modeled using a CR tomography measurement~\cite{Sheldon_AC_2016} (more details in Supplement~\cite{suppl}). $\Omega$ is the CR amplitude that can be obtained from the experimental $ZX_{90}$ pulse calibrations for each flux and gate length. Theoretical simulations agree well with experimental data (solid lines in Fig.~\hyperref[fig:twoQ_fid_flux]{\ref{fig:twoQ_fid_flux}}).

\begin{figure}
\centering
\includegraphics[width=\columnwidth]{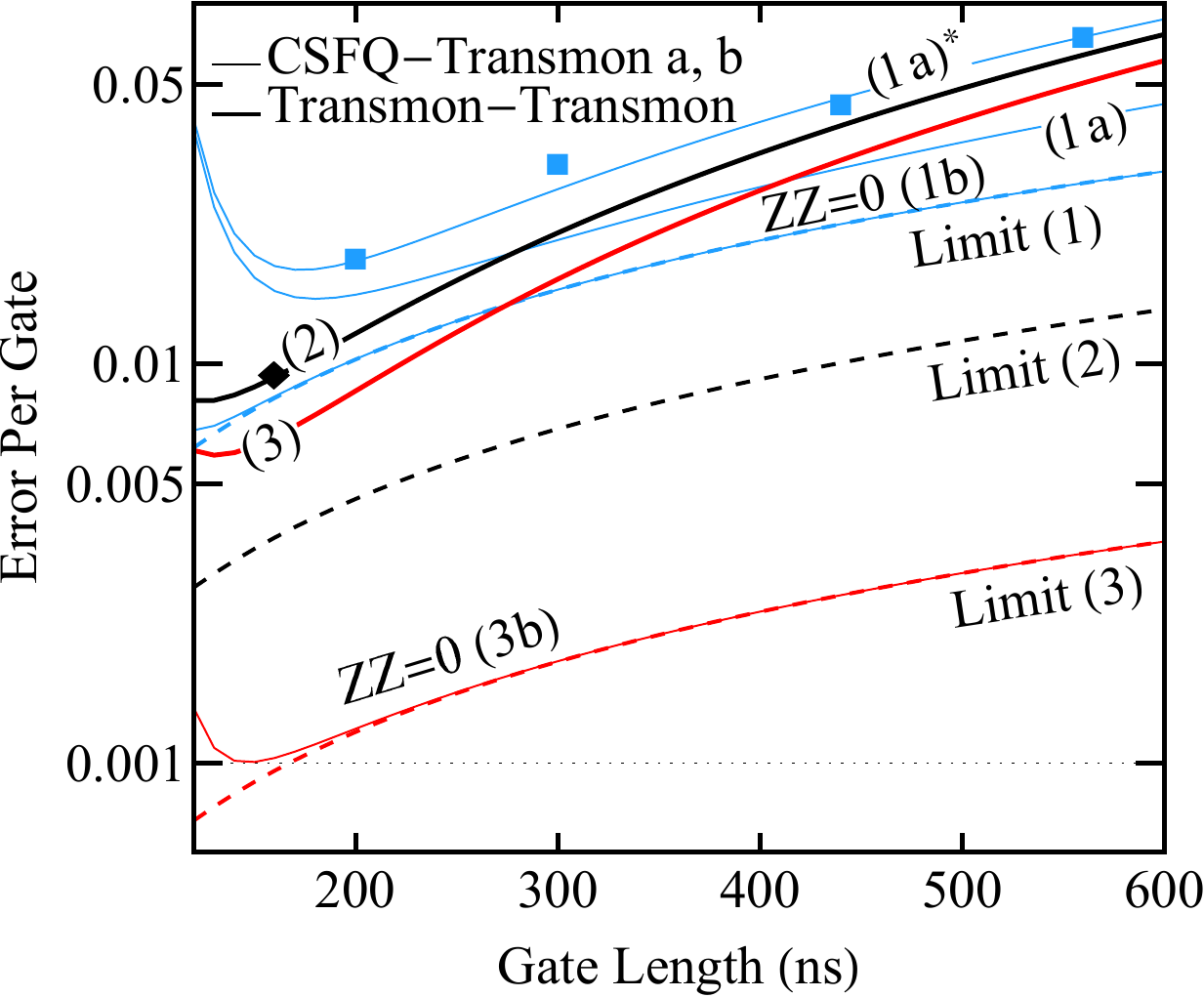}
\vspace{-0.2in}
\caption{\label{fig:twoQ_fid_predic}Experimental data and theory simulation for two-qubit gate error vs. gate length for our present CSFQ-transmon (a), static $ZZ$-free CSFQ-transmon (b), and a transmon-transmon device with non-zero $ZZ$ (thick lines). The CSFQ was placed at the flux sweet spot. Square CR pulses were used in theory simulation. Three sets of coherence times used in simulation were color-coded in blue, black, and red, and numbered by $n=\{1,2,3\}$. ``Limit (n)" represents coherence-limited gate error. Classical crosstalk is not included except $(\rm 1a)^*$. Blue squares and black diamond are experimental data points from present device and Ref.~\onlinecite{Sheldon_AC_2016}, respectively.
}
\end{figure}

Based on the success of our theoretical model in describing the measured flux- and gate-length dependence of the two-qubit gate error, we consider target parameters for a future device to achieve further reductions in gate error. In Fig.~\ref{fig:twoQ_fid_predic}, we simulate the two-qubit gate error vs. $t_g$ for three sets of coherence times in $\mu$s: ($T_1^{(1)}, T_2^{(1)}, T_1^{(2)}, T_2^{(2)}$), where the superscripts indicate the qubit, are (18, 15, 40, 45), (40, 54, 43, 67), and (200, 200, 200, 200), corresponding respectively to the present device, the two-transmon device in Ref.~\onlinecite{Sheldon_AC_2016}, and a hypothetical, but within reach, device. From the discussion above, we know that one of the most prominent advantages of a CSFQ-transmon device over a transmon-transmon device is that the static $ZZ$ interaction can be cancelled by carefully choosing qubit parameters. An idealized static $ZZ$-free device could be made by potentially keeping the CSFQ at the sweet spot, while making the transmon slightly tunable~\cite{hutchings_tunable_2017}. Such a device results in a comparable gate error (1b) for the relatively short coherence times of the present experimental device as compared to the transmon-transmon (2). For the projected longer coherence times (200 $\mu$s)~\cite{serniak_direct_2019,noauthor_cramming_2019}, the gate error (3b) of such a device subject to elimination of classical crosstalk can reach $1\times10^{-3}$. This level is inaccessible for a transmon-transmon device, even with the projected longer coherence times (3). 

While coherence-limited gate errors (dashed lines in Fig.~\hyperref[fig:twoQ_fid_predic]{\ref{fig:twoQ_fid_predic}}) decrease monotonically as gate length does, the total error reaches a minimum at an optimum gate length. This is a universal behavior, even in the absence of static $ZZ$ or classical crosstalk (e.g., (3b) in Fig.~\ref{fig:twoQ_fid_predic}), and can be explained by the dynamic $ZZ$; the $ZZ$ interaction has a static (undriven) term $\zeta$, and a dynamic (driven) term $\eta \Omega^2$, with $\eta$ being a device-dependent quantity. 
Since  $\Omega$ is larger for shorter gate length, even in the absence of a static $ZZ$ term, the dynamic part can still produce a large $ZZ$ interaction for short gate times. 

In conclusion, we have characterized the CR gate on a CSFQ-transmon device. This hybrid system with opposite anharmonicity between the qubits allows for the complete suppression of the static $ZZ$ interaction, which becomes essential for achieving a high-fidelity CR gate. Our theoretical analysis shows that suppressing the $ZZ$ interaction
is just as important as enhancing coherence times. By eliminating the spurious $ZZ$ interaction, a CSFQ-transmon gate can achieve comparable fidelities to a transmon-transmon gate despite having shorter coherence times. With longer coherence times that are not too far beyond current experimental capabilities (200 $\mu$s), two-qubit gate errors of $1\times 10^{-3}$ are feasible.

We thank David DiVincenzo, Sarah Sheldon, and Jerry Chow for helpful discussions, and acknowledge support from Intelligence Advanced Research Projects Activity (IARPA) under contract W911NF-16-0114. D.M. acknowledges support by the Army Research Office under contract W911NF-14-1-0124.

\let\oldaddcontentsline\addcontentsline
\renewcommand{\addcontentsline}[3]{}
\bibliography{CR_paper}
\let\addcontentsline\oldaddcontentsline

\widetext
\newpage

\setcounter{equation}{0}
\setcounter{figure}{0}
\setcounter{table}{0}
\setcounter{page}{1}
\makeatletter

\renewcommand\thesubfigure{(\alph{subfigure})} 
\let\theequationWithoutS\theequation 
\renewcommand\theequation{S\theequationWithoutS}
\let\thefigureWithoutS\thefigure 
\renewcommand\thefigure{S\thefigureWithoutS}
\let\thetableWithoutS\thetable 
\renewcommand\thetable{S.\thetableWithoutS}
\let\bibnumfmtWithoutS\bibnumfmt 
\renewcommand{\bibnumfmt}[1]{[S#1]}
\let\citenumfontWithoutS\citenumfont 
\renewcommand{\citenumfont}[1]{S#1}

\begin{center}
\textbf{\large Supplementary Material: Suppression of Unwanted $ZZ$ Interactions in a Hybrid Two-Qubit System}
\end{center}

{\hypersetup{linkcolor=blue}
\tableofcontents
}

\section{Device and Measurement Setup}
\begin{figure}[h]
\begin{subfloat}{\label{fig:CSFQ_SEM}}
\end{subfloat}
\begin{subfloat}{\label{fig:CSFQ_SEM_loop}}
\end{subfloat}
\includegraphics[width=0.5\columnwidth]{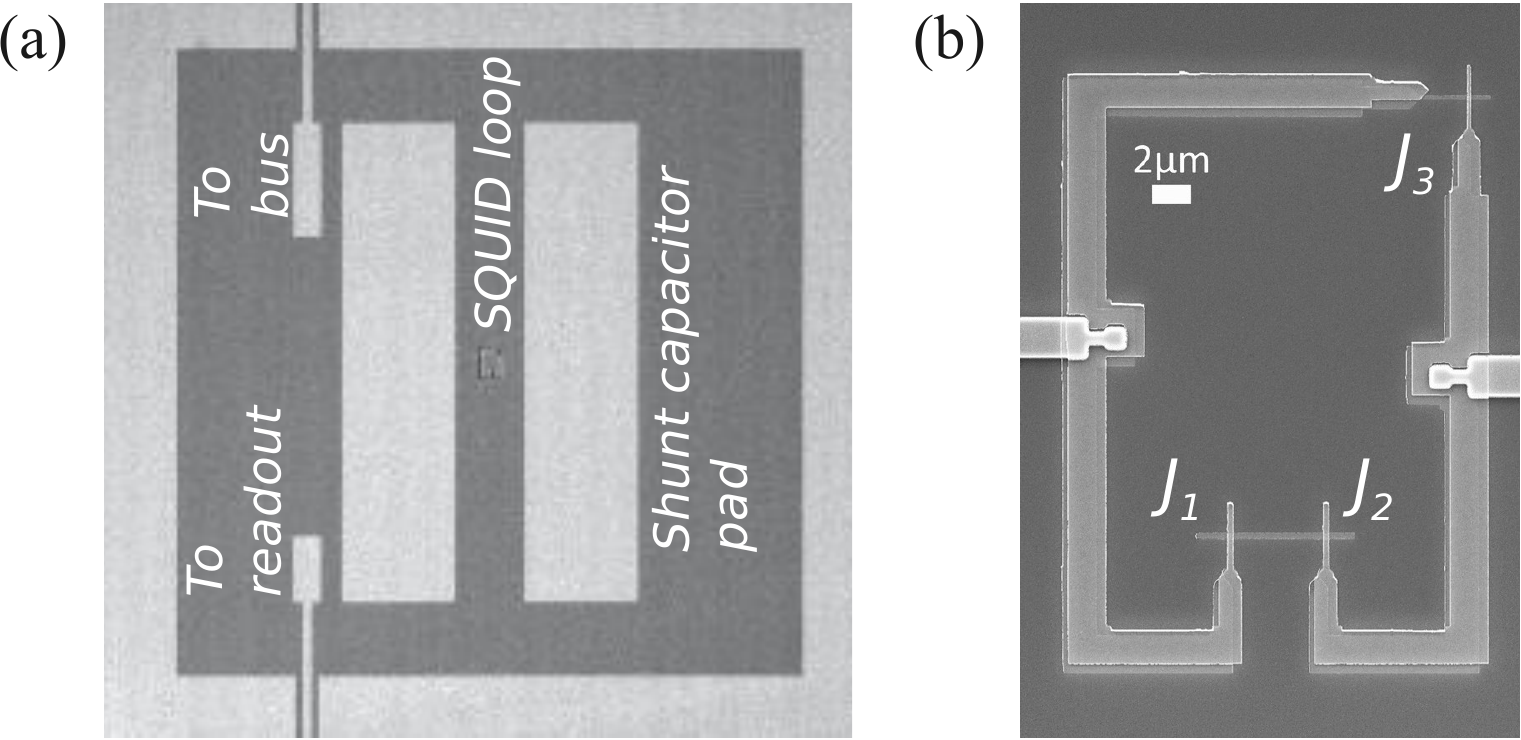}
\caption{\label{fig:SEM}SEM micrographs of CSFQ similar to the one used in this work. (a) Shunt capacitors, SQUID loop and coupling capacitors in gap within opening in chip ground plane. (b) Close-up of SQUID loop. Image of full chip of this type may be found in Ref.~\onlinecite{Sheldon_AC_2016}. $J_1$ and $J_2$ indicate two large Josephson junctions, and $J_3$ is a smaller Josephson junction.}
\end{figure}
The device was fabricated at IBM. The $4\times8$~mm$^2$ chip contains one fixed-frequency transmon, one bus resonator, one CSFQ, and readout resonators for each qubit. A photo of a similar chip appears in Ref.~\onlinecite{Supp:Sheldon_AC_2016}. We fabricated the device in a manner described in Ref.~\onlinecite{Supp:hutchings_tunable_2017,Supp:Sheldon_AC_2016}. We sputter-deposited a $\sim 200$~nm niobium film on a $730\,\mu{\rm m}$-thick silicon substrate, followed by photolithography and plasma-etch to define the microwave structures. Bus and readout resonators comprise half-wave sections of coplanar waveguide terminated by metal pads that define coupling capacitors. We formed Al/AlOx/Al tunnel junctions and the CSFQ loop using e-beam lithography, Manhattan-style double-angle shadow-evaporation~\cite{Supp:potts_cmos_2001}, and lift-off. The CSFQ contains three junctions in a $30\times20\ \mu$m$^2$ loop (Fig.~\ref{fig:SEM}). We formed the aluminum elements of the transmon and CSFQ simultaneously into identical shunting capacitors. We diced the chip, installed it into a package comprising a circuit board, a copper backing-plate, coaxial connectors and a superconducting bobbin coil. Similar packaging is described in Ref.~\onlinecite{Supp:corcoles_protecting_2011}, with the exception that the package is not potted into epoxy, but is mounted inside a light-tight magnetically shielded sample can.
\begin{figure}[t]
\begin{subfloat}{\label{fig:mw_electronics}}
\end{subfloat}
\begin{subfloat}{\label{fig:cryo_wiring}}
\end{subfloat}
\includegraphics[width=0.9\columnwidth]{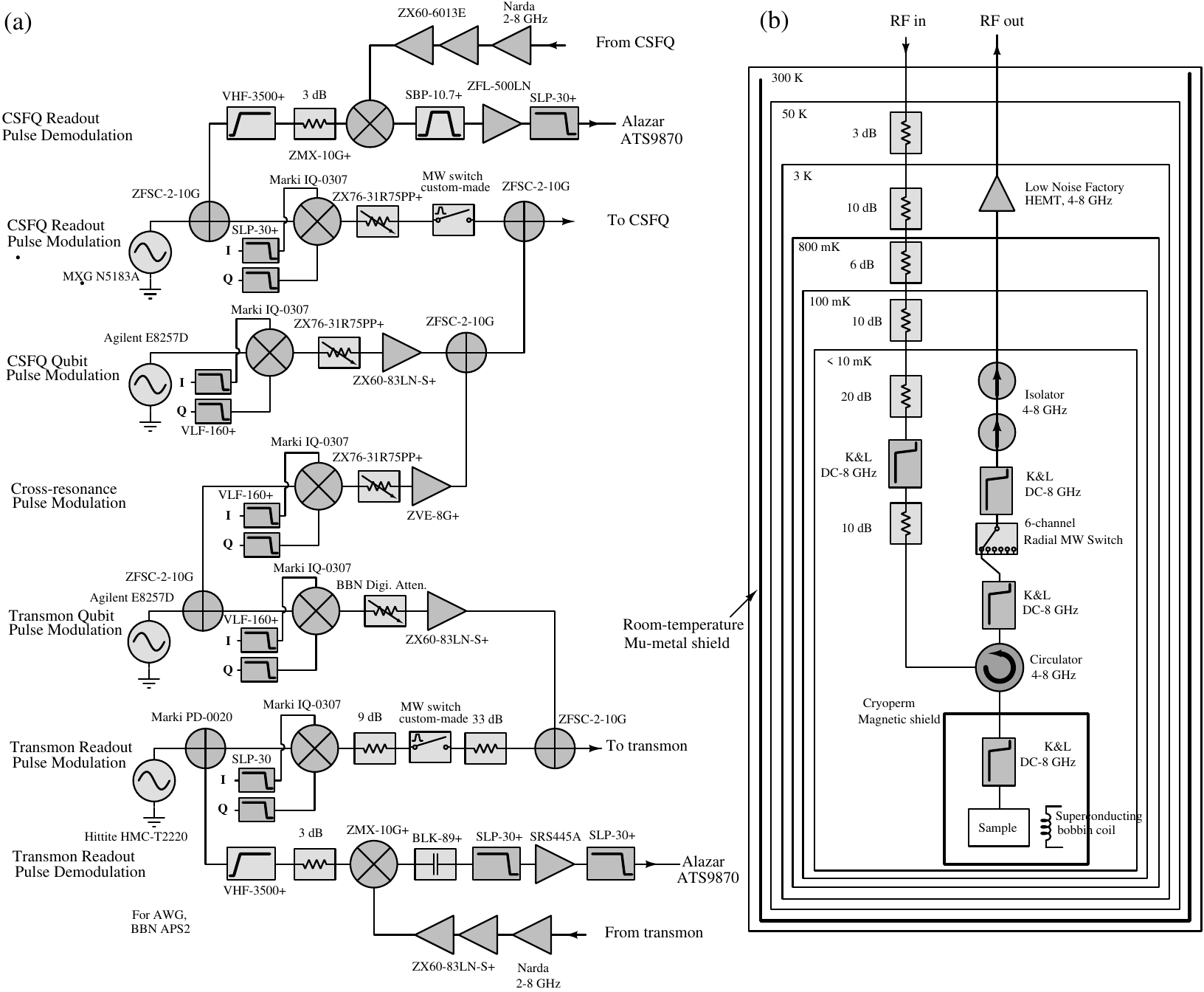}%
\caption{\label{fig:meas_setup}(a) Room-temperature microwave control electronics. (b) Cryogenic wiring for one of the two qubits. Wiring for other qubit is identical.}
\end{figure}

 The device was measured on a dry dilution refrigerator with a base temperature below 10 mK and heavily filtered cryogenic microwave lines.
We show our room-temperature microwave electronics setup in Fig.~\hyperref[fig:meas_setup]{\ref{fig:mw_electronics}} and cryogenic wiring in Fig.~\hyperref[fig:meas_setup]{\ref{fig:cryo_wiring}}. Qubit and readout pulses are created with single side band (SSB) modulation. An Arbitrary Pulse Sequencer 2 (APS2) from BBN Technologies creates I/Q signals for pulse modulation. The readout pulse is demodulated down to 10~MHz and digitized by an Alazar card. For the flux-bias, we used a battery-operated voltage source (SRS SIM928) and a 10~k$\Omega$ room-temperature standard resistor in series for a current-bias. The flux line is filtered through a $\pi$-filter at the 3~K stage and an Eccosorb filter at the mixing chamber stage before it reaches the superconducting bobbin coil inside the Cryoperm magnetic shield.

\section{Device Parameters}
In Table~\hyperref[table:device parameters]{\ref{table:device parameters}}, ~\hyperref[table:junction_params]{\ref{table:junction_params}}, ~\hyperref[table:coherence]{\ref{table:coherence}}, and~\hyperref[table:J]{\ref{table:J}}, we list device parameters.
\begin{table}[h]
\caption{\label{table:device parameters}Frequency scales on device with CSFQ at the sweet spot.}
\begin{tabular}{| l | l | l |l|}
\hline
Description & Symbol & Frequency & Method \\
\hline \hline
CSFQ bare frequency & $\omega_1^b(0)/2\pi$ & 5.0616 GHz & Calculated by solving 5 simultaneous equations~\cite{Supp:gely_nature_2018}\\ \hline
CSFQ dressed frequency & $\tilde{\omega}_1(0)/2\pi$ & 5.0511 GHz & Low-power qubit spectroscopy    \\ \hline
CSFQ anharmonicity & $\delta_1/2\pi$ & +592.7 MHz & Low- and high-power qubit spectroscopy\\ \hline
CSFQ bare readout frequency & $\omega_{h}/2\pi$ & 6.9065  GHz & High-power resonator measurement \\ \hline
CSFQ dressed readout frequency & $\tilde{\omega}_{h}/2\pi$ & 6.9074 GHz & Low-power resonator measurement  \\ \hline
CSFQ-readout coupling & $g_{hm}/2\pi$ & 34 MHz & Calculated~\cite{Supp:gely_nature_2018} \\ \hline
CSFQ-readout dispersive shift & $\chi_{hm}/2\pi$ & 550 kHz & Resonator measurement with CSFQ at $|0\rangle$ and $|1\rangle$ \\ \hline
\hline
Transmon bare frequency & $\omega_2^b(0)/2\pi$ & 5.2920 GHz & Calculated by solving 5 simultaneous equations~\cite{Supp:gely_nature_2018} \\ \hline
Transmon dressed frequency & $\tilde{\omega}_2(0)/2\pi$ & 5.2855 GHz & Low-power qubit spectroscopy  \\ \hline
Transmon anharmonicity & $\delta_2/2\pi$ & -326.6 MHz & Low- and high-power qubit spectroscopy\\ \hline
Transmon bare readout frequency & $\omega_a/2\pi$ & 6.8050 GHz & High-power resonator measurement\\ \hline
Transmon dressed readout frequency & $\tilde{\omega}_{a}/2\pi$ & 6.8059 GHz & Low-power resonator measurement \\ \hline
Transmon-readout coupling & $g_{aT}/2\pi$ & 36.2 MHz & Calculated~\cite{Supp:gely_nature_2018,Supp:koch2007charge} \\ \hline
Transmon-readout dispersive shift & $\chi_{aT}/2\pi$ & 200 kHz & Resonator measurement with transmon at $|0\rangle$ and $|1\rangle$  \\ \hline
\hline
Bus bare frequency & $\omega_r/2\pi$ & 6.3062 GHz & Calculated by solving 5 simultaneous equations~\cite{Supp:gely_nature_2018} \\ \hline
Bus dressed frequency & $\tilde{\omega}_r/2\pi$ & 6.3226 GHz & Bus cavity spectroscopy~\cite{Supp:sheldon_characterization_2017} \\ \hline
Bus-Transmon dispersive shift & $\chi_{rT}/2\pi$ & -2.2 MHz & Bus cavity spectroscopy~\cite{Supp:sheldon_characterization_2017} \\ \hline
Bus-CSFQ dispersive shift & $\chi_{rm}/2\pi$ & 5.9 MHz & Bus cavity spectroscopy~\cite{Supp:sheldon_characterization_2017}\\ \hline
Bus-CSFQ coupling & $g_{rm}/2\pi$ & 111.7 MHz & Calculated by solving 5 simultaneous equations~\cite{Supp:gely_nature_2018}\\ \hline
Bus-Transmon coupling & $g_{rT}/2\pi$ & 76.4 MHz & Calculated by solving 5 simultaneous equations~\cite{Supp:gely_nature_2018} \\  \hline
\hline
Transmon-CSFQ exchange coupling & $J_{00}/2\pi$ & 6.3  MHz & CSFQ spectroscopy and fit\\ \hline
Transmon-CSFQ direct coupling & $g_{mT}/2\pi$ & -2.7  MHz & Estimated from direct capacitance between two qubits\\ \hline
\end{tabular}
\caption{\label{table:junction_params}Junction parameters and charging energy of the CSFQ and transmon. CSFQ Josephson energy is for the larger junctions. The two transmon parameters were calculated using the measured dressed qubit frequency. Meanwhile, the three CSFQ parameters were obtained by fitting spectroscopy data of the dressed qubit frequencies, $\tilde{\omega}_1(0)/2\pi$ and $\tilde{\omega}_1(1)/2\pi$ vs. flux with 1D potential approximation~\cite{Supp:steffen_high-coherence_2010}.}
\begin{tabular}{| l | l | l |}
\hline
Description & Symbol & Value \\ \hline \hline
Transmon Josephson energy & $E_{JT}$ & 13.7 GHz  \\ \hline
Transmon charging energy &  $E_{CT}$ & 0.286 GHz  \\ \hline
\hline
CSFQ Josephson energy & $E_{Jm}$ & 123.1  GHz \\ \hline
CSFQ charging energy & $E_{Cm}$ & 0.268  GHz \\ \hline
CSFQ critical current ratio & $\alpha$ & 0.43 \\ \hline
\end{tabular}
\caption{\label{table:coherence}Coherence times for the transmon and CSFQ at the sweet spot.}
\begin{tabular}{|m{0.75cm}|m{0.75cm}|m{0.75cm}||m{0.75cm}|m{0.75cm}|m{0.75cm}|}
 \hline
 \multicolumn{3}{|c|}{Transmon} & \multicolumn{3}{|c|}{CSFQ} \\  \hline
 \hline
 $T_1$ ($\mu$s) & $T_2^*$ ($\mu$s) & $T_2$ ($\mu$s)& $T_1$ ($\mu$s) & $T_2^*$ ($\mu$s) & $T_2$ ($\mu$s)\\ \hline
 40 & 25 & 45 & 18 & 13 & 18 \\ \hline
\end{tabular}
\caption{\label{table:J}Two-photon virtual exchange coupling strength ($J_{01}$ and $J_{10}$), qubit-qubit detuning ($\Delta$), and anharmonicities ($\delta_i$) at $\Phi/\Phi_0=0.504$, where $ZZ=0$.}
\begin{tabular}{|m{1cm}|m{1cm}|m{1cm}|m{1cm}|m{1cm}|}
 \hline
 $J_{01}$ (MHz) & $J_{10}$ (MHz)& $\Delta$ (MHz) & $\delta_1$ (MHz) & $\delta_2$ (MHz)\\ \hline
 4.9 & 8.1 & 192 & 560 & -327  \\ \hline
\end{tabular}
\end{table}

\section{Theory}
\subsection{Circuit Hamiltonian}
\begin{figure}[h]
\labellist
\bfseries
\pinlabel Readout at 15 80
\pinlabel Cavity at 15 70
\pinlabel Transmon at 140 185
\pinlabel {Bus cavity} at 240 185
\pinlabel CSFQ at 360 185
\pinlabel Readout at 510 80
\pinlabel Cavity at 510 70
\endlabellist
\includegraphics[width=1\textwidth]{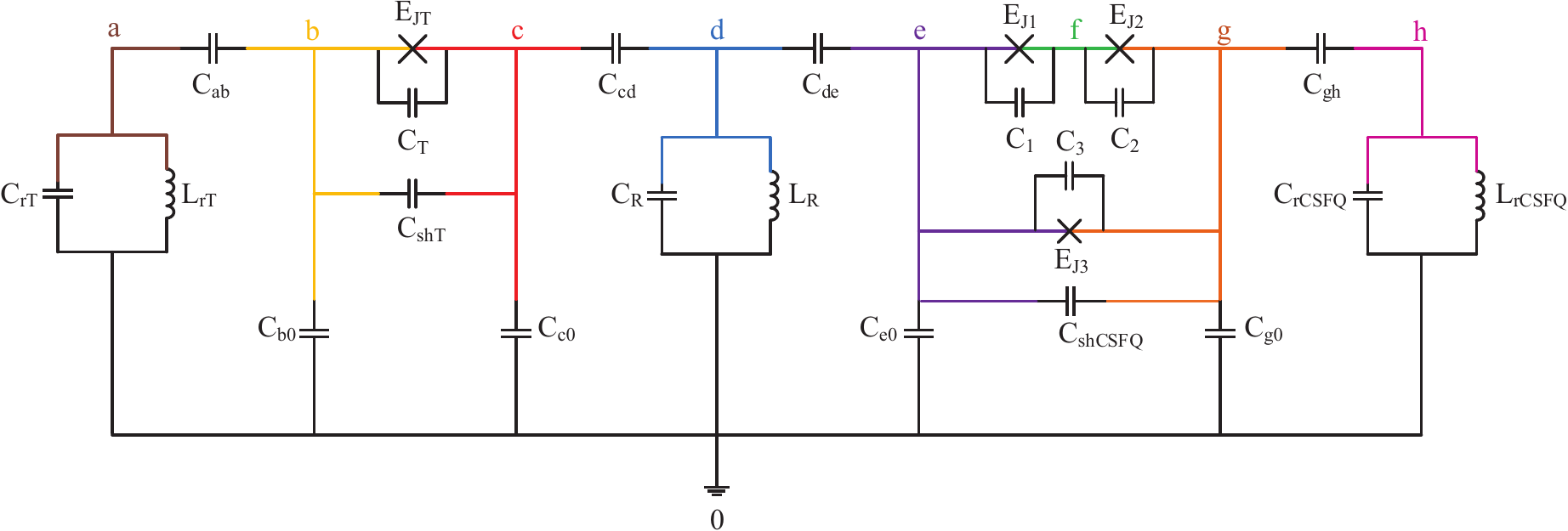} 
\caption{Circuit model for CSFQ-transmon device. Each qubit has its own readout resonator and they are coupled via a bus resonator.}
\label{figure:csfq-transmon}
\end{figure}

\begin{figure} 
\begin{subfloat}{\label{fig:potential_a}} \end{subfloat}
\begin{subfloat}{\label{fig:potential_b}} \end{subfloat}
\begin{subfloat}{\label{fig:potential_c}} \end{subfloat}
\begin{subfloat}{\label{fig:potential_d}} \end{subfloat}
\begin{subfloat}{\label{fig:potential_e}} \end{subfloat}
\begin{subfloat}{\label{fig:potential_f}} \end{subfloat}
\includegraphics[width=0.75\textwidth]{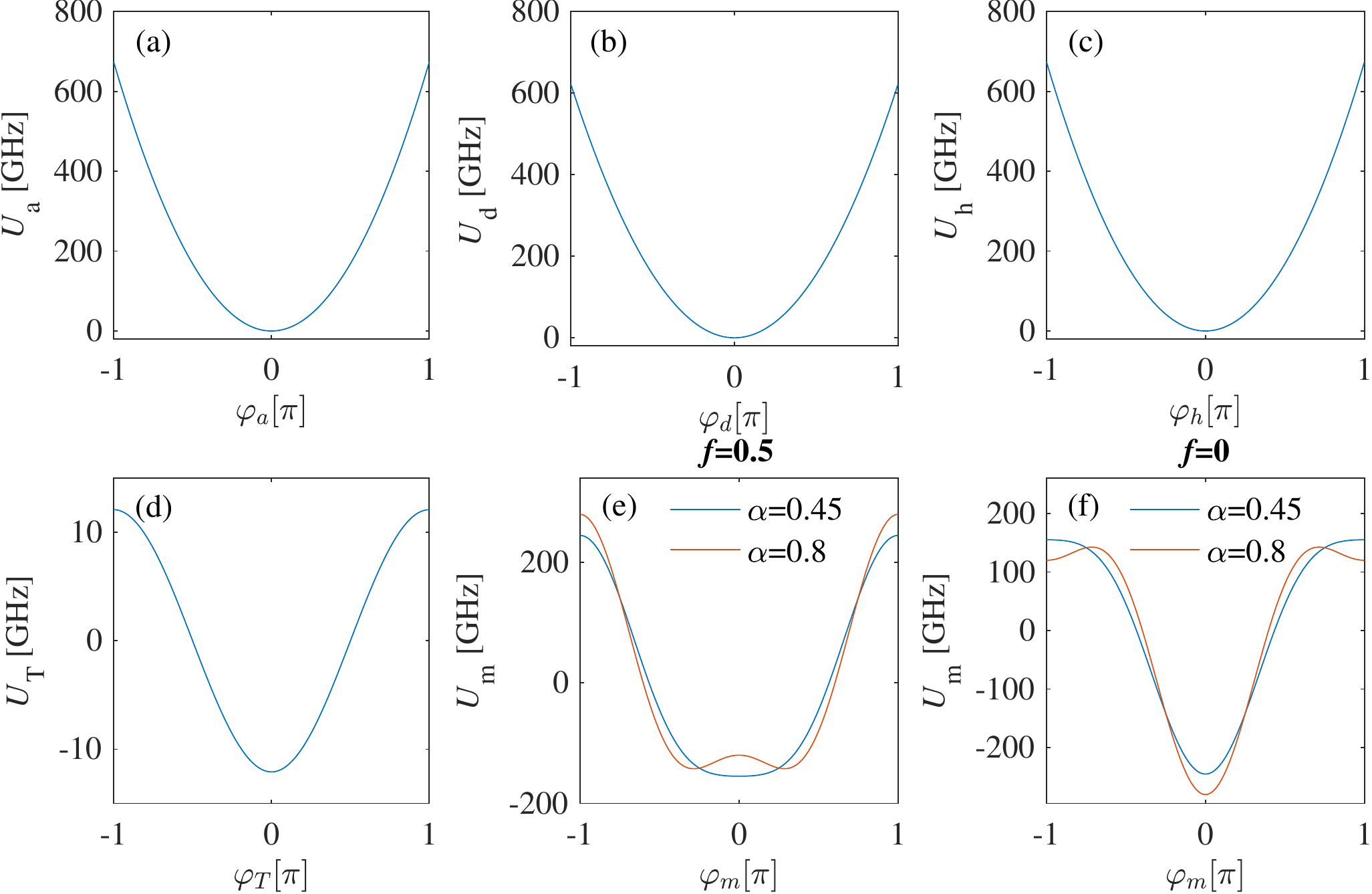}
\caption{Potential profiles. (a) Readout resonator for transmon. (b) Bus resonator. (c) Readout resonator for CSFQ. (d) Transmon. (e) CSFQ along the $\varphi_m$ direction at $f=0.5$. (f) CSFQ along the $\varphi_m$ direction at $f=0$.}
\label{allpotential}
\end{figure}

\begin{table}[h]
\begin{center}
	\begin{tabular}{|c|c|c|c|c|c|}
		\hline 
		\multicolumn{4}{|c|}{Capacitance (fF)} & \multicolumn{2}{c|}{Josephson energy (GHz)}\tabularnewline
		\hline 
		$C_{rT}$ & $452.1$& $C_{rCSFQ}$ &  $438.8$ & $E_{J1}=E_{J2}$ & $108.9$\tabularnewline
		\hline 
	 $C_{ab}$&$3.9$  &$C_{gh}$  &$3.9$ & $E_{J3}=\alpha E_{J1}$ & $46.8$\tabularnewline
		\hline 
	 	$C_{b0}$&$58$  &$C_{g0}$  & $59$  & $E_{JT}$ & $13.7$\tabularnewline
		\hline 
		$C_{shT}$&$30$ &  $C_{shCSFQ}$&$30$& $\alpha$ & $0.43$\tabularnewline
		\hline 
		 	$C_{T}$&$5$ &  $C_1=C_2$&$5$ & \multicolumn{2}{c|}{Inductance (nH)}\tabularnewline
		\hline 
	$C_{c0}$&$60$ & $C_{e0}$ &  $50.2$ & $L_{R}$ & $1.3$\tabularnewline
		\hline 
		 	$C_{cd}$ &$10 $ & $C_{de}$ & $14.5$& $L_{rT}$ & $1.2$\tabularnewline
		\hline 
			$C_R$ &$468.9$ & $C_3=\alpha C_1$ & $2.25$& $L_{rCSFQ}$ & $1.2$\tabularnewline
		\hline 
\end{tabular}
\end{center}
\caption{Circuit parameters: Capacitance, inductance and Josephson energy.}
\label{table:para}
\end{table}
We built a full-circuit Hamiltonian from a lumped-element circuit model for our CSFQ-transmon device in Fig.~\ref{figure:csfq-transmon}, and the corresponding design parameters in Table~\ref{table:para}. We write the Lagrangian $\mathcal{L}=T-U$ with $T$ being the electrostatic energy and $U$ the potential energy of the Josephson junctions, where we define $\varphi_m\equiv(\varphi_e-\varphi_g)/2$ and  $\varphi_p\equiv(\varphi_e+\varphi_g)/2-\varphi_f$ in order to simplify the Lagrangian into the following form:
\begin{equation}
\begin{split}
 \mathcal{L}&=\frac{1}{2}\left(\frac{\Phi_{0}}{2\pi}\right)^{2}\left[C_{rT}\dot{\varphi_{a}}^2+C_{ab}(\dot{\varphi_{a}}-\dot{\varphi_{b}})^{2}+C_{b0}\dot{\varphi_{b}}^{2}+C_{cd}(\dot{\varphi_{b}}-\dot{\varphi_{T}}-\dot{\varphi_{d}})^{2}\right.\\
 &+(C_{shT}+C_{T})\dot{\varphi_{T}}^{2}+C_{c0}\left(\dot{\varphi_{b}}-\dot{\varphi_{T}}\right)^{2}+C_{R}\dot{\varphi_{d}}^{2}+C_{de}(\dot{\varphi_{e}}-\dot{\varphi_{d}})^{2}+C_{e0}\dot{\varphi_{e}}^{2}\\
 &+C_{gh}\left(\dot{\varphi_{e}}-2\dot{\varphi_{m}}-\dot{\varphi_{h}}\right)^{2}+2C\left(\dot{\varphi_{m}}^{2}+\dot{\varphi_{p}}^{2}\right)+4(C_{3}+C_{shCSFQ})\dot{\varphi_{m}}^{2}\\
 &+\left.C_{g0}\left(\dot{\varphi_{e}}-2\dot{\varphi_{m}}\right)^{2}+C_{rCSFQ}\dot{\varphi_{h}}^{2}\right]+E_{JT}\cos\varphi_{T}+2E_{J}\cos{\varphi_{p}}\cos{\varphi_{m}}\\
 &+\alpha E_{J}\cos\left(2\pi f-2\varphi_{m}\right)-\left(\frac{\Phi_{0}}{2\pi}\right)^{2}\left(\frac{\varphi_{a}^{2}}{2L_{rT}}+\frac{\varphi_{d}^{2}}{2L_{R}}+\frac{\varphi_{h}^{2}}{2L_{rCSFQ}}\right),
 \end{split}
 \end{equation}
where $f=\Phi/\Phi_0$ is the normalized magnetic flux, $\Phi_0=h/2e$ the flux quantum, $h$ is Planck's constant, $e$ is the electron charge, $C\equiv C_1=C_2$, and $E_J\equiv E_{J1}=E_{J2}$. The Hamiltonian is calculated using the usual definition of $H$ as the Legendre transformation of the Lagrangian $\mathcal{L}$,
\begin{eqnarray}
 H&=&\sum_{i}\dot{\varphi_{i}}\frac{\partial \mathcal{L}}{\partial\dot{\varphi_{i}}}-\mathcal{L}=T+U,\\
 T&=&\frac{1}{2}\left(\frac{\Phi_{0}}{2\pi}\right)^{2}\vec{\dot{\varphi}}^{T}\mathbf{C}\vec{\dot{\varphi}},\\
 U&=&E_{La}\varphi_{a}^{2}+E_{Ld}\varphi_{d}^{2}+E_{Lh}\varphi_{h}^{2}+E_{JT}\cos\varphi_{T} \\ \nonumber
 &-&2E_{J}\cos{\varphi_{p}}\cos{\varphi_{m}}-\alpha E_{J}\cos\left(2\pi f-2\varphi_{m}\right),
\end{eqnarray}
where the phase vector in the circuit is defined as $\vec{\dot{\varphi}}^{T}=\left(\dot{\varphi_{b}},\dot{\varphi_{e}},\dot{\varphi_{a}},\dot{\varphi_{T}},\dot{\varphi_{d}},\dot{\varphi_{m}},\dot{\varphi_{p}},\dot{\varphi_{h}}\right)$, and the energies stored in the readout resonators for the transmon, the CSFQ, and the bus resonator are $E_{La}=\Phi_0^2/8\pi^2 L_{rT} $, $E_{LCSFQ}=\Phi_0^2/8\pi^2 L_{rCSFQ}$, and $E_{LR}=\Phi_0^2/8\pi^2 L_{R} $, respectively. Fig.~\ref{allpotential} indicates the potential energies associated with the readout resonator coupled to the transmon (a), the bus resonator (b), the readout resonator coupled to the CSFQ (c), the fixed frequency transmon (d), and the CSFQ at the sweet spot (e), and away from the sweet spot (f). In particular, the shape of the CSFQ potential in Fig.~\ref{fig:potential_e} and~\ref{fig:potential_f} shows a single well for $\alpha<0.5$ and a double well for $\alpha>0.5$. In this experiment, the ratio is designed to be less than 0.5 to be in the CSFQ regime.
Since the potential does not depend on  $\varphi_b$ and $\varphi_e$, and also the kinetic energy of $\varphi_p$ in the CSFQ is superior to its contribution in the qubit potential which leads to the fast oscillation behavior in that direction, we use standard methods to safely remove these three phases, and thus reduce the matrix size of the circuit Hamiltonian from $8\times 8$ to the following $5 \times 5$ matrix:
\begin{equation}
H=4\overrightarrow{n}^{T}\frac{e^{2}}{2\mathbf{C'}}\overrightarrow{n}+U,
\end{equation}
where $\overrightarrow{n}=(n_a, n_T, n_d, n_m, n_h)$ is the canonical term of $\overrightarrow{\varphi}$, and  
\begin{equation}
\mathbf{\mathbf{C'}}=\left(\begin{array}{ccccc}
C_{a}' & -C_{ab}C_{dT}/C_{T0}& -C_{ab}C_{cd}/C_{T0}& 0 & 0\\
 -C_{ab}C_{dT}/C_{T0} & C_{T}' & C_{a0}C_{cd}/{C_{T0}} & 0 & 0\\
-C_{ab}C_{cd}/{C_{T0}} & C_{a0}C_{cd}/{C_{T0}} & C_{r}' & -2C_{de}C_{h0}/{C_{m0}} & -C_{de}C_{gh}/{C_{m0}}\\
0 & 0 & -2C_{de}C_{h0}/{C_{T0}} & C_{m}' & 2C_{dm}C_{gh}/C_{m0}\\
0 & 0 & -C_{de}C_{gh}/{C_{m0}} & 2C_{dm}C_{gh}/{C_{m0}} & C_{h}'
\end{array}\right).
\end{equation}
The relevant capacitances are the following combinations of capacitances defined in the circuit model from Fig.~\ref{figure:csfq-transmon}:
\begin{equation}
\begin{split}
C_{T0}	&=C_{ab}+C_{b0}+C_{c0}+C_{cd},C_{h0}	=C_{g0}+C_{gh}\\
C_{m0}	&=C_{de}+C_{e0}+C_{g0}+C_{gh}, C_{dm}=C_{de}+C_{e0}\\
C_{dT}	&=C_{cd}+C_{c0},C_{a0}=C_{ab}+C_{b0}\\
C_{T}'	&=C_{dT}+C_{shT}+C_{T}-{C_{dT}^{2}}/{C_{T0}}\\
C_{m}'	&=2C+4(C_{3}+C_{shCSFQ})-4C_{h0}^{2}/C_{m0}+4C_{h0}\\
C_{r}'	&=-C_{cd}^{2}/C_{T0}+C_{cd}+C_{de}+C_{r}-{C_{de}^{2}}/{C_{m0}}\\
C_{a}'	&=-C_{ab}^{2}/C_{T0}+C_{ab}+C_{rT}\\
C_{h}'    &=-C_{gh}^{2}/{C_{m0}}+C_{gh}+C_{rCSFQ}.
\end{split}
\end{equation}
Analytical expressions for the transmon frequency and anharmonicity can be obtained using the systematic perturbation theory to large orders~\cite{Supp:gely_nature_2018}. Similarly, the quantization of the CSFQ requires that we define the following operators in the Fock space~\cite{Supp:in_preparation}:
\begin{equation}
\varphi_m=\xi (m+m^{\dagger}),\  n_m=\frac{i}{2\xi}(m^{\dagger}-m),
\end{equation}
where $\xi$ is a device-dependent parameter. Fig.~\ref{freandanharm} shows the theoretical flux dependence of the bare frequency and anharmonicity in our experimental device. After quantizing the circuit, we simplify its Hamiltonian by taking it to a rotating frame and applying the Rotating Wave Approximation (RWA), which results in:
\begin{figure}
\begin{subfloat}{\label{fig:sub1}}
\centering
\includegraphics[width=0.6\columnwidth]{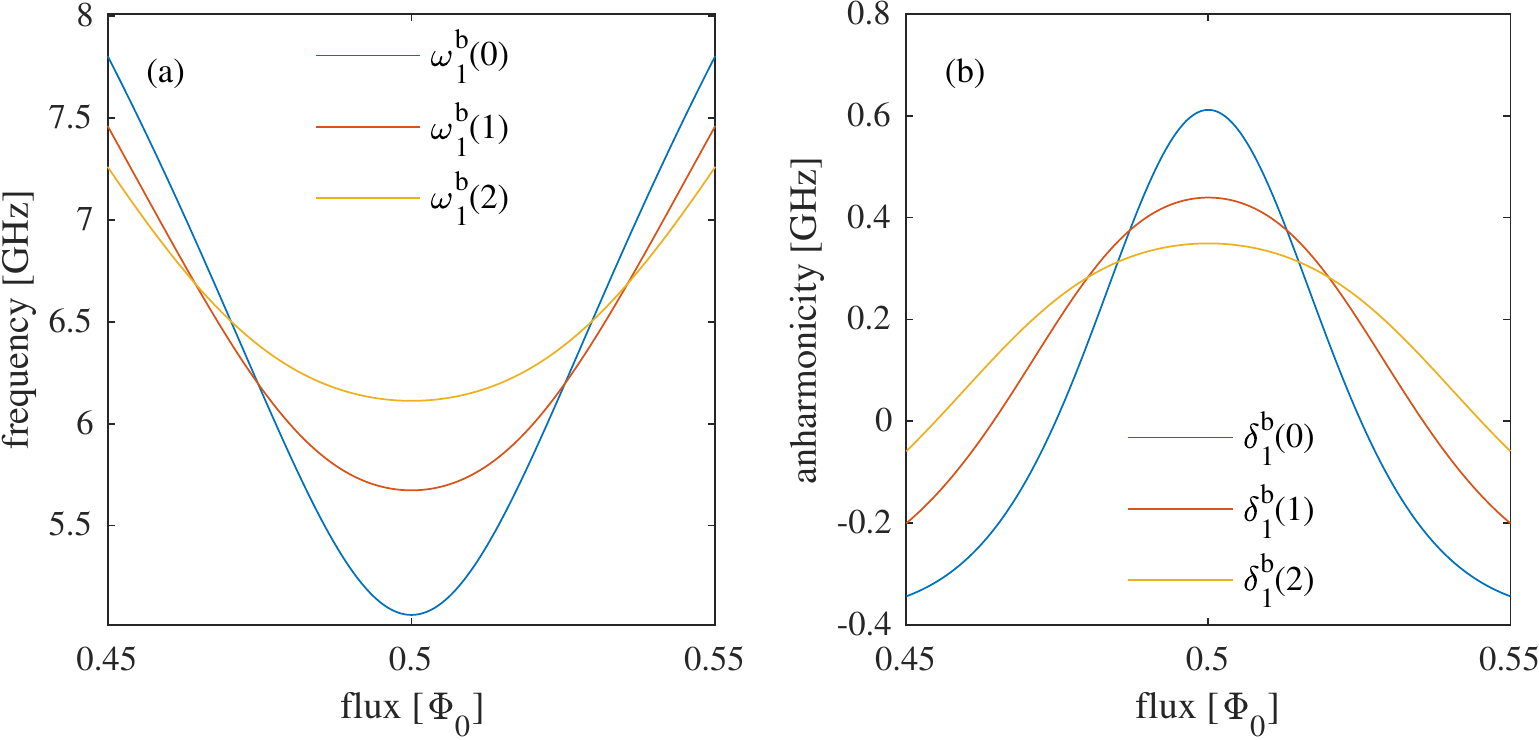}
\end{subfloat}
\begin{subfloat}{\label{fig:sub2}}
\end{subfloat}
\caption{Bare CSFQ frequency and anharmonicity versus flux. (a) Bare transition frequency $\omega_1^b(n)$ between levels $n$ and $n+1$. (b) Bare CSFQ anharmonicity $\delta_1^b(n)=\omega_1^b(n+1)-\omega_1^b(n)$.}
\label{freandanharm}
\end{figure}

\begin{equation}
\begin{split}
H_{\rm circuit}&=\omega_{a}a^{\dagger}a+g_{aT}\left(a^{\dagger}T+aT^{\dagger}\right)+\omega_{h}h^{\dagger}h+g_{hm}\left(h^{\dagger}m+hm^{\dagger}\right)\\
&+\omega_{r}r^{\dagger}r+\sum_j\omega_T(j)\left|j\right>\left<j\right|+\sum_k\omega_m(k)\left| k \right>\left< k\right|+g_{rm}\left(r^{\dagger}m+rm^{\dagger}\right)\\
&+g_{rT}\left(r^{\dagger}T+rT^{\dagger}\right)+g_{mT}\left(m^{\dagger}T+mT^{\dagger}\right),
\end{split}
\end{equation}
where $a$ and $h$ represent the readout resonators, $r$ is the bus resonator, and we use $m=\sum_k\sqrt{k+1}\left|k\right>\left<k+1\right|$ and $T=\sum_j\sqrt{j+1}\left|j\right>\left<j+1\right|$ for the CSFQ and transmon, respectively. We use the notation $\omega_1^b(j)$ to denote the bare transition frequency between the energy levels, $j+1$ and $j$ in the transmon; similarly, $\omega_2^b(k)$, the bare transition frequency between the energy levels, $k+1$ and $k$ in the CSFQ. The relationships between the various coupling strengths $g_{ij}$ and the relevant capacitances are given by the following expressions:
\begin{equation}
\begin{split}
g_{hm}	&\sim -\frac{2C_{gh}C_{dm}}{\left(C_{gh}+C_{rCSFQ}\right)\left(C_{gs}C_{m0}-4C_{h0}^{2}\right)} \\
g_{rm}&\sim\frac{2C_{de}C_{h0}}{C_{cder}\left(4C_{h0}^{2}-C_{gs}C_{m0}\right)}\\
g_{aT}&\sim\frac{C_{ab}C_{dT}}{\left(C_{ab}+C_{rT}\right)\left(C_{gT}C_{T0}-C_{a0}^{2}\right)} \\
g_{rT}&\sim -\frac{C_{cd}C_{a0}}{C_{cder}\left(C_{dT}^{2}-C_{gT}C_{T0}\right)}\\
g_{mT}&\sim-\frac{2C_{cd}C_{de}C_{a0}C_{h0}}{C_{cder}\left(C_{gT}C_{T0}-C_{a0}^{2}\right)\left(C_{gs}C_{m0}-4C_{h0}^{2}\right)},
\end{split}
\end{equation}
where $C_{gs}=2C_{0}+4C_{g0}+4C_{gh}+4(C_{3}+C_{shCSFQ})$,  $C_{gT}=C_{cd}+C_{c0}+C_{shT}+C_{T}$, and $C_{cder}=C_{cd}+C_{de}+C_{R}$. 
In the limit that the qubit-resonator detuning is much larger than the coupling between the qubits and resonators, we can use the Schrieffer-Wolff transformation to simplify the Hamiltonian. Here we first eliminate the readout resonators and then the bus, and obtain the multilevel version of the qubit-qubit effective Hamiltonian:
\begin{equation}\label{eq:twoQ_Hamil}
\begin{split}
H_{qr}&=H_r+H_q=\tilde{\omega}_{r}r^{\dagger}r+\sum_{q=1,2}\sum_{n_q}{\omega}_q(n_q)\left|n_q\right>\left<n_q\right|\\
    &+\sqrt{(n_1+1)(n_2+1)}J_{n_1,n_2}\left(\left|n_1+1,n_2\right>\left<n_1,n_2+1\right|+\left|n_1,n_2+1\right>\left<n_1+1,n_2\right|\right),
\end{split}
\end{equation}
where the dressed bus frequency is $\tilde{\omega}_r=\omega_r+\Sigma_q \chi_{n_q}^q\left|n_q\right\rangle \left\langle n_q\right|$, and $\chi$ is the dispersive shift of the resonator frequency, which can be solved using Eq. (9) in Ref.~\onlinecite{Supp:gely_nature_2018}, $J_{n_1, n_2}$ is the two-photon virtual coupling rate defined as  $J_{j,k}=J^{\text{dir}}+J_{j,k}^{\text{indir}}$ with the direct coupling being $J^{\text{dir}}=g_{mT}$, and the indirect coupling $J_{j,k}^{\text{indir}}$:
\begin{eqnarray}
J_{j,k}^{\text{indir}}&=&-\frac{g_{rm}^{k,k+1}g_{rT}^{j,j+1}}{2}\left(\frac{1}{\Delta_m(k)}+\frac{1}{\Delta_T(j)}+\frac{1}{\Sigma_m(k)}+\frac{1}{\Sigma_T(j)}\right) \\ \Delta_m(k)&=&\omega_r-\omega_m(k) \\ \Delta_T(j)&=&\omega_r-\omega_T(j)\\
\Sigma_m(k)&=&\omega_r+\omega_m(k) \\ 
\Sigma_T(j)&=&\omega_r+\omega_T(j).\\
\end{eqnarray}
In the limit of $|\Delta|\gg J$, the Hamiltonian [Eq.~\eqref{eq:twoQ_Hamil}] can be diagonalized into the Hamiltonian in the dressed frame, using a unitary operator $U$:
\begin{equation}
\tilde{H_q}=U^{\dagger}H_q U=\sum_{q=1,2}\sum_{n_q}\tilde{{\omega}}_q(n_q)\left|n_q\right>\left<n_q\right|.
\end{equation}
The dressed qubit frequencies, anharmonicity, bare bus frequency, coupling strength, and two-photon exchange rate are presented in Table~\ref{table:device parameters}, where we define $\tilde{\omega}_q\equiv\tilde{\omega}_q(0)$ and $g_{\alpha\beta}\equiv g_{\alpha\beta}^{01}$.


\subsection{Cross-Resonance Gate}
A cross-resonance gate is enabled by driving the control qubit at the frequency of the target qubit and this allows for a two-qubit entangling gate between two qubits, where additional single-qubit rotations can implement a CNOT operation. In the dressed frame, the CR driving Hamiltonian is,
\begin{equation}
\tilde{H_{\rm d}}=U^{\dagger}H_{\rm d}U=\Omega\cos(\omega_{\rm d} t) \sum_{n_1} U^{\dagger}(\left|n_1\right\rangle \left\langle n_1+1\right|+\left|n_1+1\right\rangle \left\langle n_1\right|)U.
\end{equation}
Moving into the rotating frame by RWA,
\begin{equation}
H_r=R^{\dagger}(\tilde{H}+\tilde{H_{\rm d}}) R-i R^{\dagger}R,
\end{equation}
where $R=\sum_n\exp(-i\omega_{\rm d} t \hat{n})\left|n\right\rangle \left\langle n\right|$. For our device, we consider the total number of excitations to be limited to 4, therefore we consider the states \{00, 01, 10, 11, 02, 20, 03, 12, 21, 30, 04, 13, 22, 31, 40\}. Next, we block-diagonalize it into two individual qubit blocks and a block for all higher excited levels -- 2$\times$2, 2$\times$2, and 11$\times$11 -- to decouple the higher levels from the computational subspace under the principle of least action~\cite{Supp:Cederbaum_1989}. This method aims to find a unitary operator $T$, which is closest to the identity operation. The least action unitary operator $T$ that satisfies $H_{BD}=T^{\dagger} H_r T$ is given by~\cite{Supp:Sheldon_AC_2016, Supp:magesan2018effective}
\begin{equation}
T=X X_{BD}^{\dagger}X_P^{-\frac{1}{2}},
\end{equation}
where $X$ is the nonsingular eigenvector matrix of $H_r$, $X_{BD}$ is the block-diagonal matrix of $X$, and $X_P=X_{BD}X_{BD}^{\dagger}$. Finally, the driven Hamiltonian in the computational subspace can be written as 
\begin{equation}
H_{\rm CR}=\alpha_{ZI}\frac{ZI}{2}+\alpha_{IX}\frac{IX}{2}+\alpha_{ZX}\frac{ZX}{2}+\zeta(\Omega) \frac{ZZ}{4}.
\end{equation}
The CR gate is accompanied with some unwanted interactions such as $ZZ$, $IX$, and $ZI$. The latter two can be cancelled out by echoed CR sequences~\cite{Supp:Corcoles_RB_2013}, while the $ZX$ term remains and results in the rotation of the target qubit on the Bloch sphere.
On top of the static $ZZ$ interaction $\zeta$, which solely comes from the contribution of higher excitations in the qubit-qubit interaction,  the CR drive with the amplitude $\Omega$ introduces an additional $ZZ$ interaction that depends quadratically on $\Omega$. The two together produce the total $ZZ$ interaction, $\zeta(\Omega)=\zeta(0) + \eta\Omega^2$, where $\zeta(0)$ is the static $ZZ$ interaction, and $\eta\Omega^2$ is what we refer to as the dynamic $ZZ$ interaction. In this manuscript, we look into schemes for eliminating the static $ZZ$ interaction, while in Ref.~\onlinecite{Supp:in_preparation}, we demonstrate a scheme for eliminating the total ZZ interaction -- $\zeta(\Omega)$.

\subsection{Classical Crosstalk}
\begin{figure}[b]
\includegraphics[width=0.4\linewidth]{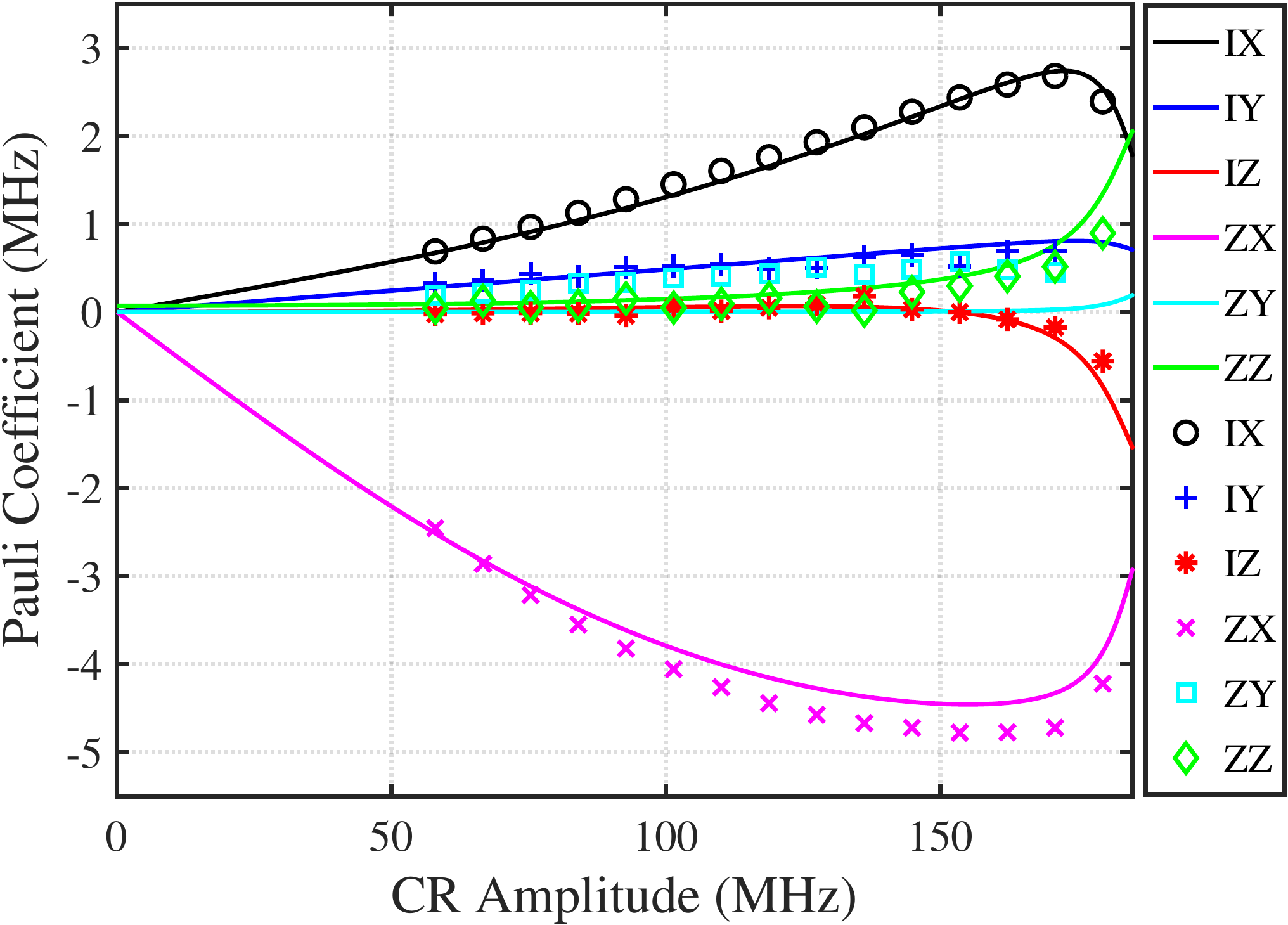}
\caption{Pauli coefficients vs. CR amplitude at the sweet spot for 200-ns gate: experimental tomographic measurements (points) and theoretical curves (solid lines). The parameters used for simulation are $R=0.0125$, $\phi_0=\pi$ and $\phi_1=\pi+0.4$ in Eq.~\eqref{eq:cc_driving_Hamil}.}
\label{paulicoeff}
\end{figure}
In the presence of classical crosstalk, the normal driving Hamiltonian can be modified to have a drive term on the target qubit,
\begin{equation}\label{eq:cc_driving_Hamil}
\begin{split}
H_{\rm d}^{\rm ct}&=\Omega\cos(\omega_{\rm d} t+\phi_0)\sum_{n_1}(\left|n_1\right\rangle \left\langle n_1+1\right|+\left|n_1+1\right\rangle \left\langle n_1\right|)\\
&+R\Omega\cos(\omega_{\rm d} t+\phi_1)
\sum_{n_2}(\left| n_2 \right\rangle \left\langle n_2+1\right|+\left|n_2+1\right\rangle \left\langle n_2\right|),
\end{split}
\end{equation}
where $R$ is a scaling factor for classical crosstalk amplitude, and depends on both two-qubit gate length and flux. $\phi_0$ is the phase of the CR drive to the control qubit and $\phi_1$ is the phase lag on the target qubit. 
When the Hamiltonian is taken to the dressed frame and block diagonalized, one can find the terms $IY$ and $ZY$ in the effective driving Hamiltonian below:
\begin{equation}
H_{\rm CR}^{\rm{ct}}=\beta_{ZI}\frac{ZI}{2}+\beta_{ZX}\frac{ZX}{2}+\beta_{ZY}\frac{ZY}{2}+\beta_{IX}\frac{IX}{2}+\beta_{IY}\frac{IY}{2}+\beta_{ZZ} \frac{ZZ}{4},
\end{equation}
where the Pauli coefficients are
\begin{equation}\label{eq:Pauli_coeff}
\begin{split}
\beta_{ZX}&\approx(B_f\Omega+C_f\Omega^3)\cos\phi_0\\
\beta_{ZY}&\approx(B_f\Omega+C_f\Omega^3)\sin\phi_0\\
\beta_{IX}&\approx(D_f\Omega+E_f\Omega^3)\cos\phi_0+RK_f\Omega\cos\phi_1\\
\beta_{IY}&\approx(D_f\Omega+E_f\Omega^3)\sin\phi_0+RK_f\Omega\sin\phi_1\\
\beta_{ZI}&\approx\alpha_{ZI}\\
\beta_{ZZ}&\approx\zeta(\Omega),
\end{split}
\end{equation}
where $B_f$, $C_f$, $D_f$, $E_f$ and $K_f$ are flux-dependent quantities that can be evaluated numerically. Here we show one example of active cancellation measurement using CR tomography~\cite{Supp:Sheldon_AC_2016}, where the driving phase was calibrated as $\phi_0=\pi$ and $\phi_1=\pi+0.4$. All experimental and theoretical Pauli coefficients at the sweet spot are plotted in Fig.~\ref{paulicoeff}. One can see that the unwanted $ZY$ vanishes in the device, and the $IY$ component can be classically removed by applying the compensation tone with a negative phase to the target qubit (see Ref.~\onlinecite{Supp:Sheldon_AC_2016} for more details),although this was not done in our measurements. 

\subsection{Echoed CR Frequency}
\begin{figure}[b]
    \centering
    \includegraphics[width=0.5\linewidth]{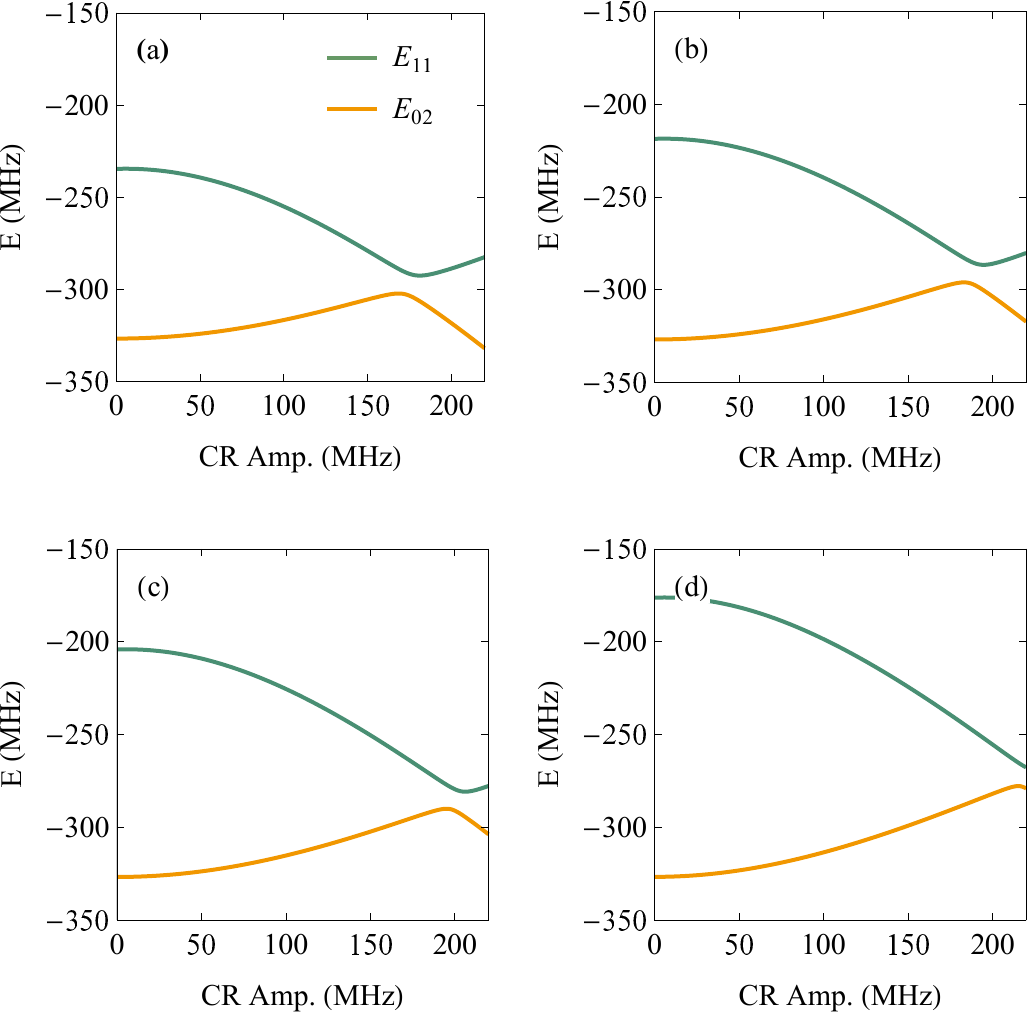}
    \caption{Energy levels, $E_{11}$ and $E_{02}$, evolving with CR amplitude for different flux bias points: (a) $f=0.5$, (b) $f=0.5026$, (c) $f=0.5036$, and (d) $f=0.505$.}
    \label{energy1102}
\end{figure}
After eliminating all unwanted components, the CR gate will effectively behave like a two-qubit gate corresponding to $ZX_{\theta}$~\cite{Supp:divincenzo2013quantum}:
\begin{equation}
    ZX_{\theta}=\exp{\left[-i \theta \left(ZX/2\right)\right]}=\left(\begin{array}{cccc}
    \cos(\theta/2)&-i\sin(\theta/2)&0&0\\
    -i\sin(\theta/2)&\cos(\theta/2)&0&0\\
    0&0&\cos(\theta/2)&i\sin(\theta/2)\\
    0&0&i\sin(\theta/2)&\cos(\theta/2)\\
    \end{array}\right).
\end{equation}
For our particular entangling gate, we choose $\theta=-\pi/2$. Following the echoed CR gate shown in Fig.~3 inset in the main text, one can find that in the presence of a $ZZ$ interaction as well as all other unwanted terms, the frequency of the echoed CR oscillation, $f_{\rm ECR}$, can be determined from the following relation (see Ref.~\onlinecite{Supp:in_preparation} for more details):
\begin{equation}\label{eq:fCR}
    2\pi f_{\rm ECR}=\sqrt{\left(\beta_{ZX}+\beta_{IX}\right)^2+\left(\beta_{ZY}+\beta_{IY}\right)^2+\left(\beta_{ZZ}/2\right)^2}+\sqrt{\left(\beta_{ZX}-\beta_{IX}\right)^2+\left(\beta_{ZY}-\beta_{IY}\right)^2+\left(\beta_{ZZ}/2\right)^2}.
\end{equation}
If both the classical crosstalk and $ZZ$ interaction are eliminated, Eq.~\eqref{eq:fCR} reduces to 2$\beta_{ZX}$. As shown in Fig.~3 in the main text, there is an upper limit to the echoed CR frequency, which arises when the two energy levels, $E_{11}$ and $E_{02}$, get closer together for increasing CR drive amplitude and leakage begins to play a role. The energy eigenvalues of the $E_{11}$ and $E_{02}$ levels in the rotating frame are shown in Fig.~\ref{energy1102}. The maximum value of $f_{\rm ECR}$ occurs at the CR amplitude where the $E_{11}$ and $E_{02}$ levels have an anti-crossing.

For a $ZX_{90}$ rotation, the length of each CR pulse in the echoed CR pulse sequence and frequency of the echoed CR gate satisfy $(2\pi f_{\rm ECR})\tau = \pi/2$, where $\tau$ is the length of CR pulses in the echoed CR pulse when the CR pulses are assumed to be \textit{square} pulses. In practice, we use Gaussian flat-top CR pulses with Gaussian rising and falling edges, where $\tau_0$ is defined to be the length of the flat-top part of each CR pulse. Due to the finite rise and fall time, we have $\tau>\tau_0$, e.g., for $\tau_0=0$, $\tau$ is non-zero. The gate length is defined as $t_g = 2\tau_0+120$~ns with average 20-ns rise/fall time of each of the two CR pulses and 40-ns $\pi$-pulses, leading to the 160-ns constant term at the end of the $t_g$ expression. Fig. 3 in the main text shows that for a weak driving regime, $f_{\rm ECR} \approx \gamma(f)\Omega$ with a flux-dependent coefficient $\gamma(f)$, e.g., $\gamma(0.5) \approx 0.1$. The exact flux-dependent $\gamma(f)$ can be found from Eq.~\eqref{eq:fCR}. Eliminating $f_{\rm ECR}$ from the expression earlier in this paragraph, we obtain the following expression for the CR amplitude for a $ZX_{90}$ gate with CR pulse length $\tau$ and flux bias $f$:
\begin{equation}\label{eq:Omega}
\Omega(f, \tau)=1/[4\gamma(f)\tau].
\end{equation}

\subsection{Two-qubit Gate Error}
We simulate an echoed CR pulse sequence for implementing a $ZX_{90}$ gate in order to compute the two-qubit error per gate by considering the density matrix starting in the ground state in the Pauli basis. Here, the $ZZ$ interaction is a global error, and for each time step we apply corresponding operators and decoherence terms. The total map is, 
\begin{equation}
\rho_{t}=\Lambda_{T1,T2,Q1}\circ\Lambda_{T1,T2,Q2} \circ\Lambda_{ZZ}\circ \Lambda_{XI} \circ \Lambda_{\rm CR-}\circ \Lambda_{XI}\circ \Lambda_{\rm CR+}[\rho_i],
\end{equation}
where each map is defined by,
\begin{equation*}
\begin{split}
\Lambda_{ZZ}[\rho]&=U_{ZZ} \cdot \rho \cdot U_{ZZ}^{\dagger}\\
\Lambda_{XI}[\rho]&=XI \cdot \rho \cdot XI \\
\Lambda_{\rm CR\pm}[\rho]&=U_{\rm CR\pm} \cdot \rho \cdot U_{\rm CR\pm}^{\dagger}\\
\Lambda_{T_1,T_2}[\rho]&=\frac{1-e^{-t_g/T_2}}{2} {\rm{Z}} \cdot \rho \cdot {\rm{Z}} +\frac{1+e^{-t_g/T_2}}{2} \rho \\
&+\frac{1-e^{-t_g/T_1}}{2}\left|0\right>\left<1\right| \cdot \rho \cdot \left|1\right>\left<0\right|-\frac{1-e^{-t_g/T_1}}{2}\left|1\right>\left<1\right| \cdot \rho \cdot \left|1\right>\left<1\right|,
\end{split}
\end{equation*}
where the unitary operators, $U_{ZZ}$ and $U_{\rm CR\pm}$, are defined as $U_{ZZ}=e^{-i 2 \pi \zeta(\Omega)t_g ZZ/4}$ , and $U_{\rm CR\pm}=e^{-i 2 \pi \tau H_{\rm CR}^{\rm{ct}}(\pm\Omega)}$ , where $t_g$ is the gate length as defined at the end of the previous section.

\begin{figure}[h]
\includegraphics[width=0.35\linewidth]{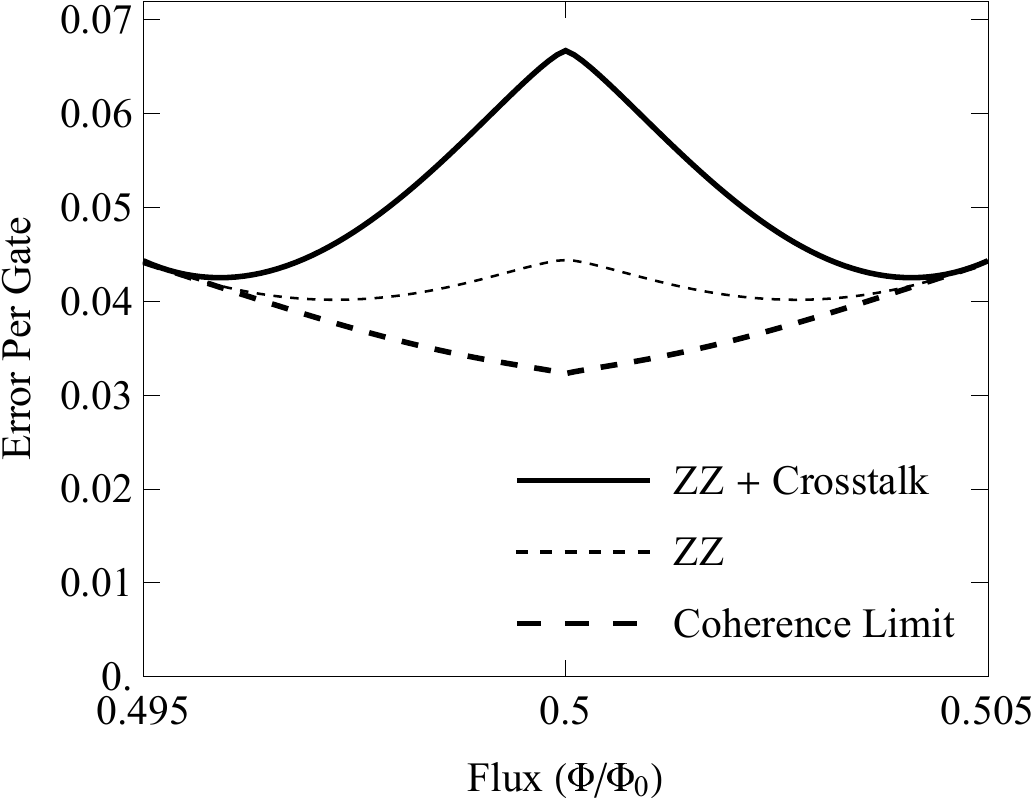}
\caption{Error per gate for $t_g=560$ ns with three different cases of error sources. Thick dashed line corresponds to coherence-limited gate error, which sets the lower bound for error per gate. Thin dashed line shows the gate error when $ZZ$ contribution is added in the simulation. Solid line shows the case where both $ZZ$ and classical crosstalk are included.}
\label{fidelity560}
\end{figure}
\begin{figure}
\makebox[\linewidth]{
\begin{subfloat}{\label{fig:coh_limit_gate_error}}
\labellist
\bfseries
\pinlabel (a) at 10 400
\endlabellist
\includegraphics[width=0.31\linewidth]{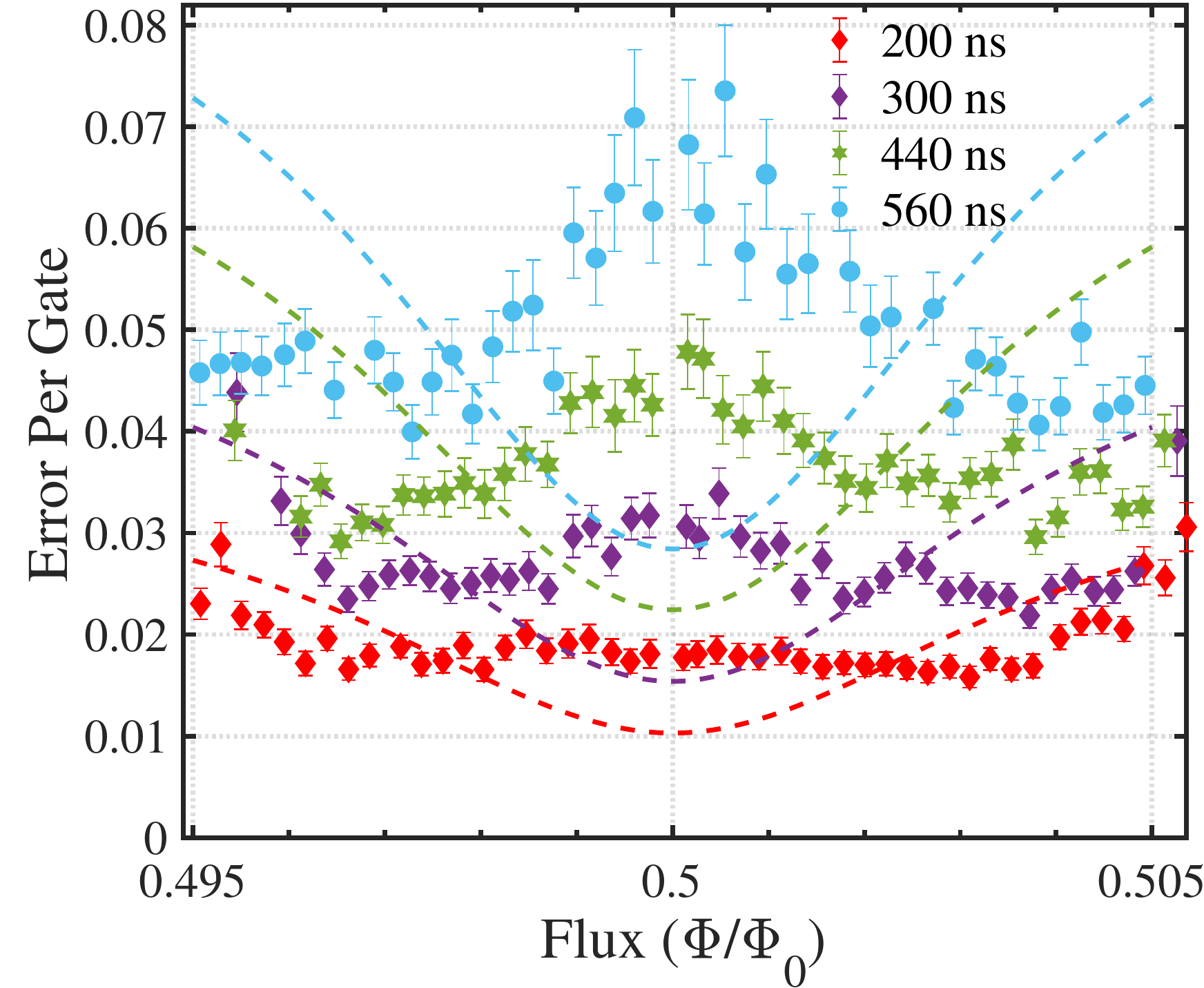}
\end{subfloat}
\labellist
\bfseries
\pinlabel (b) at 10 400
\endlabellist
\begin{subfloat}{\label{fig:Gamma_vs_DPhi}}
\includegraphics[width=0.31\linewidth]{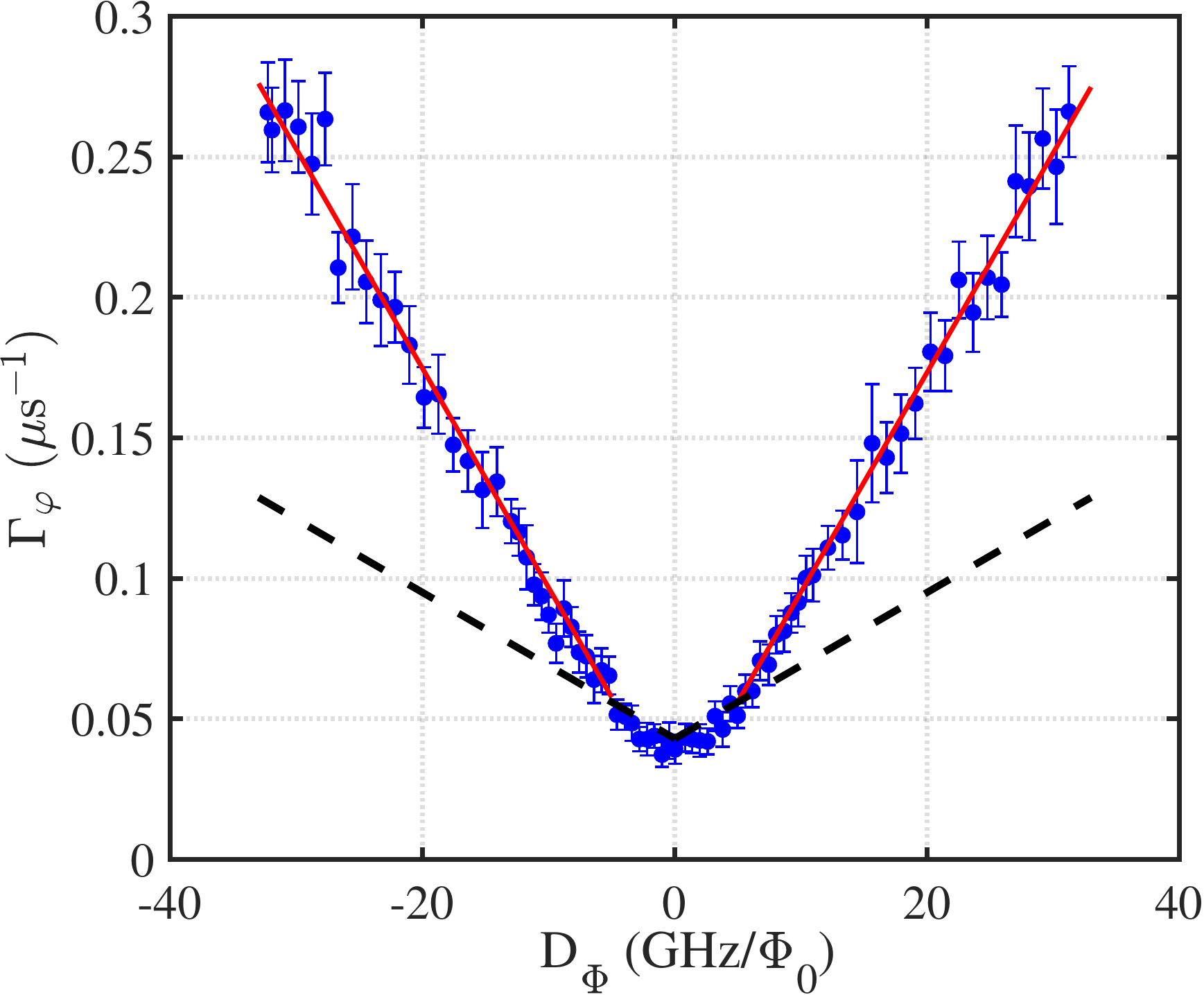}
\end{subfloat}
\begin{subfloat}{\label{fig:T2_flux_modified}}
\labellist
\bfseries
\pinlabel (c) at 10 400
\endlabellist
\includegraphics[width=0.3\linewidth]{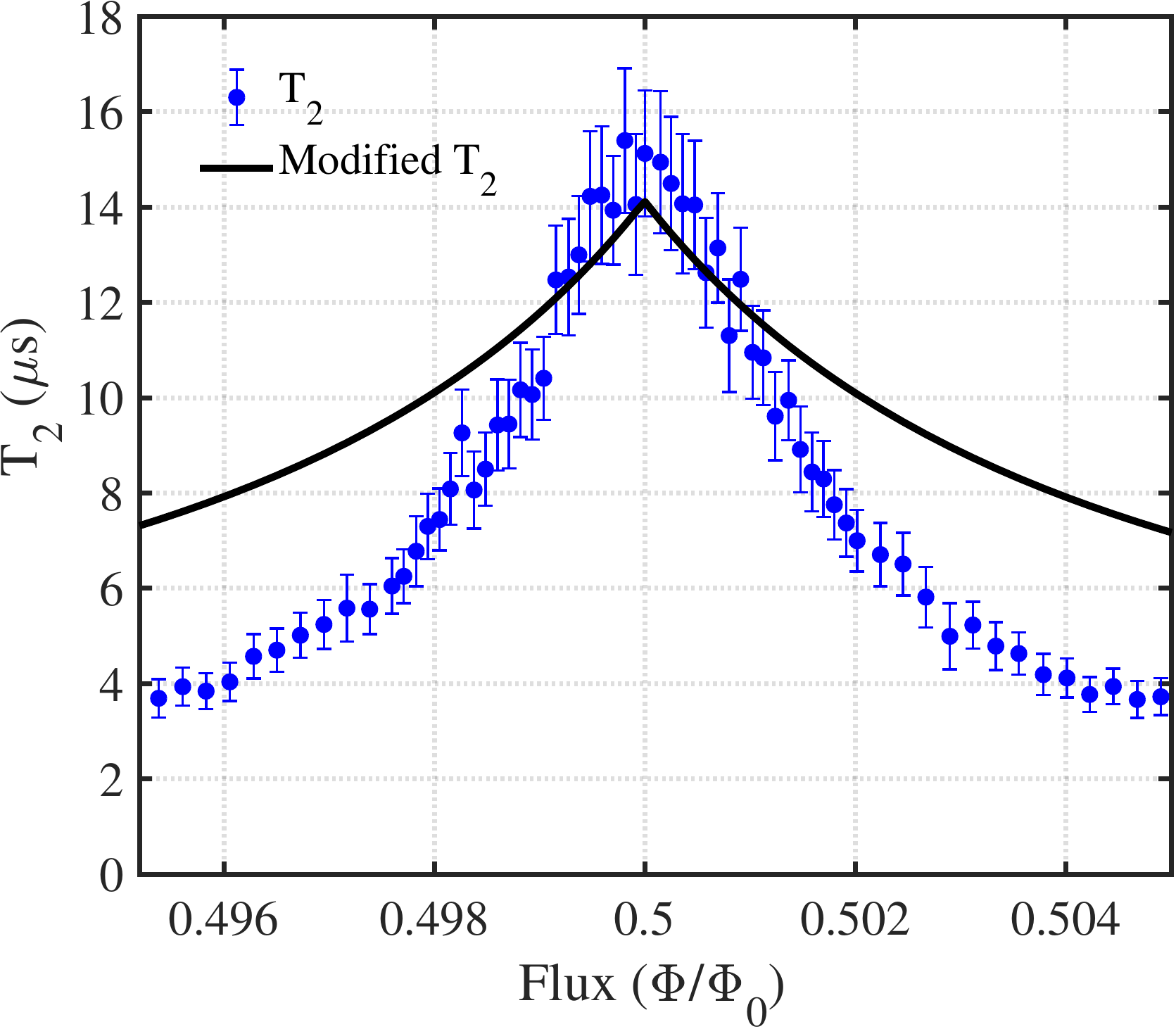}
\end{subfloat}
}
\caption{(a) Two-qubit gate error vs. flux. Dashed lines represent coherence-limited gate errors computed with measured $T_1$ and $T_2$ values, with $T_2$ extracted from a Hahn-echo measurement. (b) Pure dephasing rate $\Gamma_\varphi$ of the CSFQ vs. qubit frequency gradient $D_\Phi$. Red solid line is a linear fit to the linear portion of data. Black dashed line is the modified pure dephasing rate to account for the reduction in gate errors measured with randomized benchmarking when dephasing is dominated by $1/f$ noise, as discussed in text. (c) Hahn-echo $T_2$ vs. flux and the effective $T_2$ calculated using our modified dephasing model. }
\label{theoT2}
\end{figure}

To explore the impact of unwanted interactions, we plot the simulated gate error of echoed CR pulses at different CSFQ flux points for $t_g=560$~ns in Fig.~\ref{fidelity560}. We plot the flux-dependent gate error for three cases: coherence-limited error only (thick dashed line), $ZZ$ but no classical crosstalk included (thin dashed line), and both $ZZ$ and classical crosstalk included (solid line). Both classical crosstalk and $ZZ$ interaction add to the gate error the most at the flux sweet spot since, despite the longest coherence time, the $ZZ$ interaction and classical crosstalk are maximal at this point. Away from the sweet spot, all unwanted interactions become suppressed, and therefore the gate fidelity approaches its coherence limit. For this plot, we take all parameters from the present device; for the static $ZZ$ interaction, we use experimental data shown in Fig. 2 in the main text; for the flux dependence of $T_2$ in the CSFQ, we use the effective $T_2$ shown in Fig.~\ref{fig:T2_flux_modified} and discussed below. 
To simulate the gate error for different flux bias points, we must model the appropriate flux-dependent dephasing of the CSFQ. Simply taking $T_2$ to be the value obtained by a Ramsey measurement with a single echo refocusing pulse is insufficient and overestimates the gate errors away from the sweet spot [see Fig.~\ref{fig:coh_limit_gate_error}]. Such overestimation suggests that the effective $T_2$ used to compute the two-qubit gate error must be longer than what is measured with the standard Hahn-echo protocol using  a  single  echo  ($Y_{180}$) pulse. Single-qubit randomized benchmarking in the presence of $1/f$ dephasing noise was studied theoretically in Ref.~\onlinecite{Supp:epstein_investigating_2014}. In this case, the fidelity decay for a simulated randomized benchmarking measurement with $1/f$ dephasing noise was significantly slower compared to a simulated sequence with non-$1/f$ dephasing noise for the same $T^*_2$ Ramsey decay time. The authors of Ref.~\onlinecite{Supp:epstein_investigating_2014} suggest that this behavior may be due to the depolarizing effect from twirling the $1/f$ noise with random Clifford gates for a randomized benchmarking sequence. In the case of flux-tunable qubits, such as CSFQs or transmons, flux noise with a $1/f$ spectrum is typically the dominant contribution to dephasing for bias points away from a sweet spot~\cite{Supp:yoshihara_decoherence_2006, Supp:hutchings_tunable_2017, yan_flux_2016}.

The effect of $1/f$ flux noise on two-qubit gate errors with flux-tunable qubits has been discussed previously~\cite{Supp:rol_fast_2019, Supp:hong_demonstration_2020}. However, we are not aware of prior studies of the effect of $1/f$ noise on two-qubit gate errors measured with randomized benchmarking. Thus, we follow an approach to determine an effective $T_2$ for our gate error measurements with randomized benchmarking in the presence of $1/f$ flux noise. We first characterize the pure dephasing of our CSFQ as a function of flux with a standard echo sequence at each flux bias point. Near the sweet spot, the echo data follows an exponential decay, while away from the sweet spot the decay is Gaussian, characteristic of dephasing due to $1/f$ noise~\cite{Supp:yoshihara_decoherence_2006}. To find the pure dephasing rate $\Gamma_\varphi$, we fit the Hahn-echo decay data at each flux bias point with a Gaussian decay model, $\Gamma_\varphi = A+B\exp(-t/(2T_1)-(t/T_\varphi)^2)$~\cite{Supp:yoshihara_decoherence_2006,Supp:braumller2020characterizing}, where $T_1=18~\mu\rm s$ is fixed based on a separate relaxation measurement and $\{A,B, T_\varphi\}$ are fitting parameters. The dephasing rate is calculated by $\Gamma_\varphi=1/T_\varphi$ and plotted as a function of the derivative of the qubit transition frequency with respect to flux $D_\Phi=\partial f_{01}/\partial \Phi$. In Fig.~\ref{fig:Gamma_vs_DPhi}, $\Gamma_\varphi$ is linear with respect to $D_\Phi$ over nearly the full range of $D_\Phi$, which is consistent with dephasing dominated with flux noise. The offset of $\Gamma_{\phi}$ at $D_{\Phi}=0$ is due to a flux-independent non-$1/f$ noise source, such as photon-number fluctuations in the CSFQ readout resonator~\cite{Supp:sears_photon_2012}; the slight deviation from linearity near $D_{\Phi}=0$ is likely due to some other unknown high-frequency noise source and we do not account for it in our model. Using the slope from a linear fit to the data away from $D_{\Phi}=0$, we are able to apply the expression $\Gamma_\varphi = 2\pi\sqrt{A_\Phi\ln 2}D_\Phi$~
\cite{yoshihara_decoherence_2006} to extract a flux noise amplitude at 1~Hz, $A^{1/2}_\Phi=1.5\ \mu\Phi_0$. In order to model the reduced effective dephasing for randomized benchmarking measurements of gate errors in the presence of $1/f$ dephasing noise, we reduce this slope by a factor of 2.7, such that the calculated coherence-limited gate error from the resulting effective $T_2$ remains lower than the experimental gate error data. At the same time, we leave the offset of $\Gamma_{\varphi}$ at $D_{\Phi}=0$ unchanged, as the effects of dephasing from the non-$1/f$ noise at the sweet spot are unlikely to be mitigated by the application of random Clifford sequences. The black dashed line in Fig.~\ref{fig:Gamma_vs_DPhi} shows the modeled pure dephasing rate, $\Gamma_\varphi=(0.00288\,{\rm m}\Phi_0)D_\Phi+0.039\ \mu\rm s^{-1}$ for $D_\Phi \ge 0$. Using this modified $\Gamma_\varphi$, we calculate the effective $T_2$ vs. flux for RB as shown in Fig.~\ref{fig:T2_flux_modified}. This approach to accounting for gate error measurements with RB sequences in the presence of $1/f$ noise results in calculated coherence-limited gate error vs. flux curves that agree reasonably well with our experimental data for 4 different two-qubit gate lengths in Fig. 4 of the main text.

Classical crosstalk is caused by stray microwave coupling to the target qubit when driving the control qubit for a cross resonance gate. Thus, such crosstalk induces additional on-resonance rotation of the target qubit on the Bloch sphere during a CR drive. We can model the effect of this crosstalk in our experimental system as an additional term in Eq.~\eqref{eq:cc_driving_Hamil}: a classical crosstalk with amplitude $R(f, \tau)\Omega(f,\tau)$, where $R(f,\tau)$ is a scaling factor for modeling flux- and gate-length dependent classical crosstalk. Our experiment on the CSFQ-transmon device revealed that for shorter CR gates, e.g., $\sim 200$~ns, the corresponding classical crosstalk on the target qubit is much less significant than for the longer gate lengths. This implies that $R(f, \tau)$ is an increasing function of the gate length. For simplicity, we consider that $R(f,\tau)$ is separable, and we find $R(f, \tau)\approx\alpha(f)\tau^{2/3}$ to give good agreement with the experimental data over the full flux range. The nonlinearity with respect to $\tau$ was introduced since otherwise the $IY$ Pauli coefficient at the corresponding CR amplitude for short gates is consistently larger than the result from measured CR tomography, for example, in Fig.~\ref{paulicoeff}. In order to extract the flux-dependence of $\alpha(f)$, we performed separate CR tomography measurements~\cite{Supp:Sheldon_AC_2016} for active cancellation of classical crosstalk. We performed such an active cancellation experiment for a fixed gate length at three different flux points and found that away from the sweet spot, the classical crosstalk amplitude followed a nearly linear decrease with respect to flux, described by the fitting function: $R=(0.07-40|f-0.5|^{1.2})\tau^{2/3}$. For the simulation of two-qubit gate error at the flux sweet spot, we used $R=(0.0123,0.0220,0.0322,0.0383)$ and $\Omega/2\pi=\{70,30,17,12\}$~MHz for the gate length $t_g=(200, 300, 440, 560)$~ns, respectively.

\begin{figure}[h]
\centering
\includegraphics[width=0.9\linewidth]{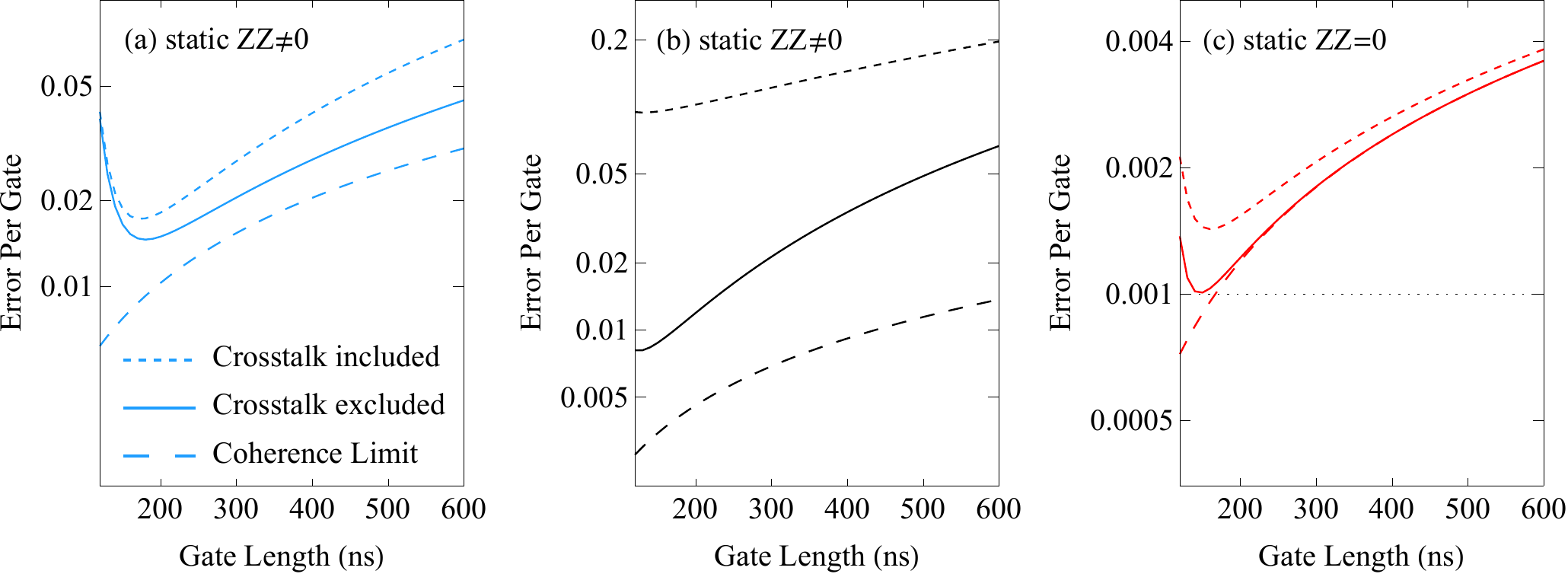}
\caption{Two-qubit gate error for three sets of coherence times.  ($T_1^{(1)}, T_2^{(1)}, T_1^{(2)}, T_2^{(2)}$), where the superscripts indicate the qubit, are (18, 15, 40, 45) for the present device, (40, 54, 43, 67) for the two-transmon device in Ref.~\onlinecite{Supp:Sheldon_AC_2016}, and (200, 200, 200, 200) for an ideal CSFQ-transmon device (all times in $\mu$s). (a) Present CSFQ-transmon device. (b) Transmon-transmon device. (c) Ideal CSFQ-transmon device. Note that all three figures share the same legend as in (a). CSFQ is assumed to be at the sweet spot in the simulation. }
\label{fidelitycoherence}
\end{figure}

To make a comparison of the gate error between a CSFQ-transmon hybrid device and an all-transmon device in Fig. 5 of the main text, we considered a state-of-the-art transmon-transmon device~\cite{Supp:Sheldon_AC_2016} and also an ideal CSFQ-transmon device. For the ideal CSFQ-transmon device, we set the static $ZZ=0$ at the sweet spot for which we use the current circuit parameters and only change the Josephson energy $E_J$. By changing the gate length, we determine the gate fidelity $F$ and plot the gate error, defined as $1-F$, in Fig.~\ref{fidelitycoherence}. The analysis shows that the error rate of $1\times10^{-3}$ can be achieved in a CSFQ-transmon device with no static $ZZ$ term, no classical crosstalk, and enhanced coherence ($T_1,\ T_2 = 200\,\mu{\rm s}$). The corresponding coherence times are listed in Table~\ref{coherence limit}.
\begin{table}[h]
    \centering
    \begin{tabular}{|c|c|c|c|c|c|c|c|c|c|}
\hline 
    Device & $T_{1}^{(1)}$& $T_{2}^{(1)}$& $T_{1}^{(2)}$ &$T_{2}^{(2)}$&$\tilde{\omega}_1$ &$\tilde{\omega}_2$ &$\delta_1$&$\delta_2$&$\eta$ \tabularnewline
& $\mu$s&$\mu$s& $\mu$s &$\mu$s& GHz& GHz& MHz& MHz&1/MHz\tabularnewline
\hline
      Present CSFQ-transmon &18&15&40&45&5.051&5.286&$+593$&$-327$&$6.0\times 10^{-5}$ \tabularnewline
\hline 
      transmon-transmon&40&54&43&67&5.114&4.914&$-330$&$-330$ & $1.6\times 10^{-5}$\tabularnewline
\hline 
        Ideal CSFQ-transmon &200&200&200&200&5.094&5.286&$+593$&$-327$&$8 \times 10^{-6}$\tabularnewline
\hline 
    \end{tabular}
    \caption{Coherence time, transition frequency, anharmonicity and nonlinear $ZZ$ interaction rate for the current device, transmon-transmon and ideal CSFQ-transmon device, respectively.}
    \label{coherence limit}
\end{table}
\section{Experiment}
\subsection{Characterizing Static $ZZ$ Interaction}
The static $ZZ$ interaction was measured at different flux points by the JAZZ (Joint Amplification of $ZZ$) protocol~\cite{Supp:garbow_bilinear_1982, Supp:takita_experimental_2017}. This measurement protocol involves a Ramsey measurement on one qubit with an echo $\pi$-pulse inserted to both qubits. The pulse sequence is executed twice -- once with the qubit that is not manipulated by the Ramsey measurement in the ground state, then again in the excited state. 
The frequency difference between the two resultant Ramsey fringes then corresponds to the static $ZZ$ strength. It is necessary to vary the phase of the second $\pi/2$-pulse to observe fringes, since the $\pi/2$-pulses are on resonance for each qubit. The oscillation frequency of the fringes, and hence the extracted $ZZ$ strength, is independent of the choice of qubit for the Ramsey sequence. Because the transmon has better coherence, we chose it for the Ramsey measurement.
 \subsection{CSFQ Coherence Versus Flux}
\begin{figure}[h]
\begin{subfloat}{\label{fig:T1}}
\labellist
\bfseries
\pinlabel (a) at 15 400
\endlabellist
\includegraphics[width=0.32\columnwidth]{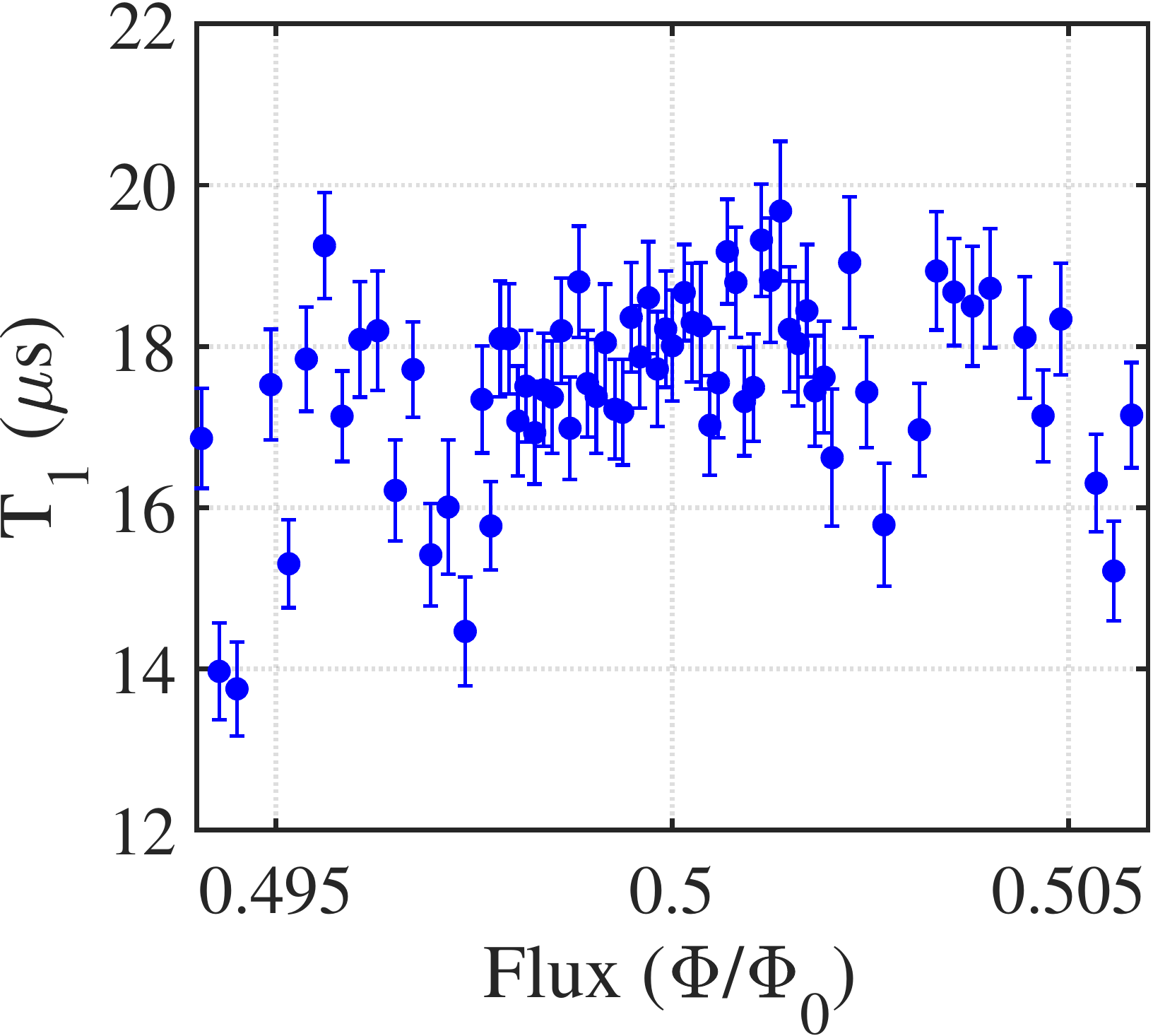}
\end{subfloat}
\begin{subfloat}{\label{fig:T2_Ramsey}}
\labellist
\bfseries
\pinlabel (b) at 15 400
\endlabellist
\includegraphics[width=0.325\columnwidth]{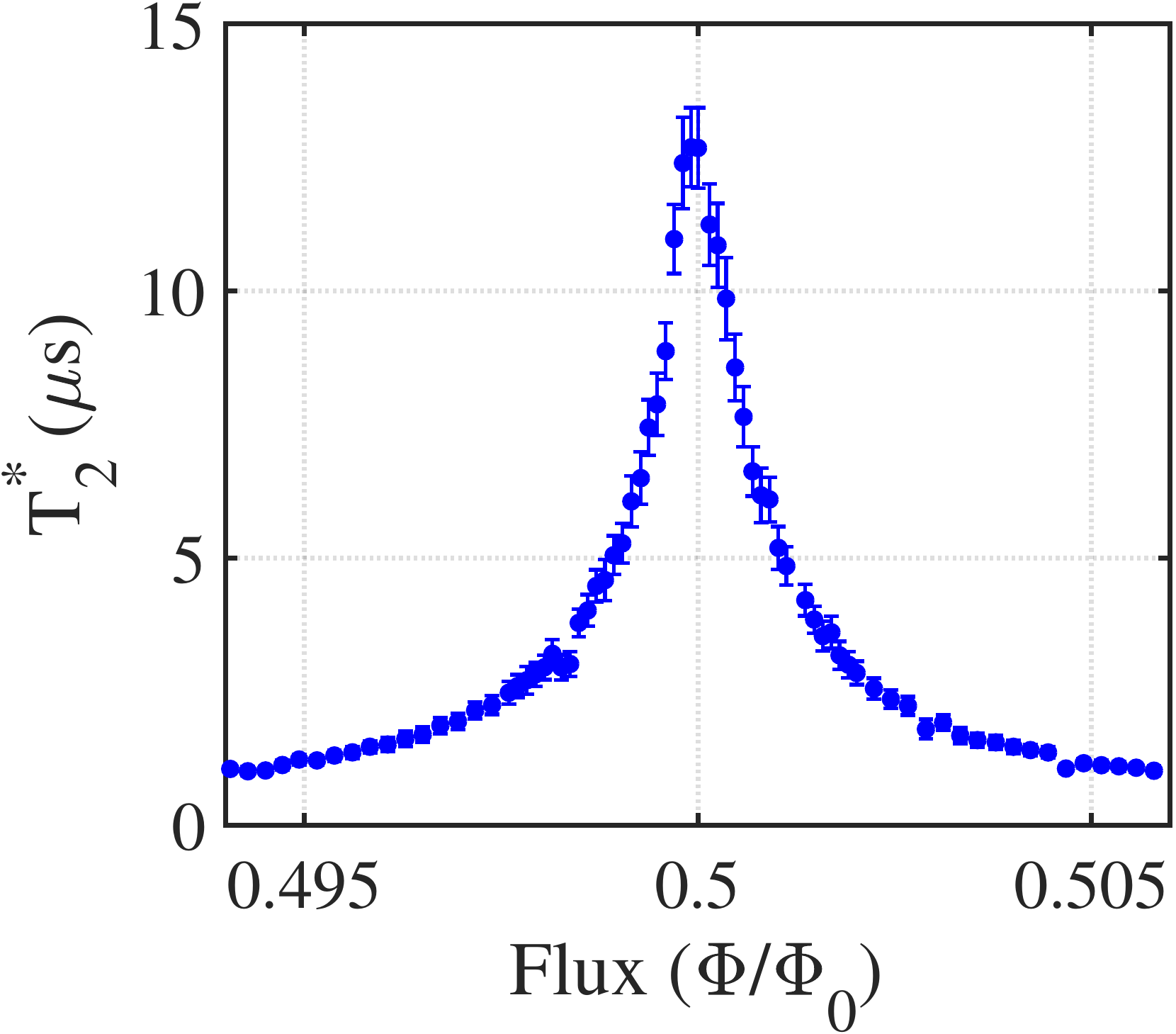}
\end{subfloat}
\begin{subfloat}{\label{fig:T2_Echo}}
\labellist
\bfseries
\pinlabel (c) at 15 400
\endlabellist
\includegraphics[width=0.325\columnwidth]{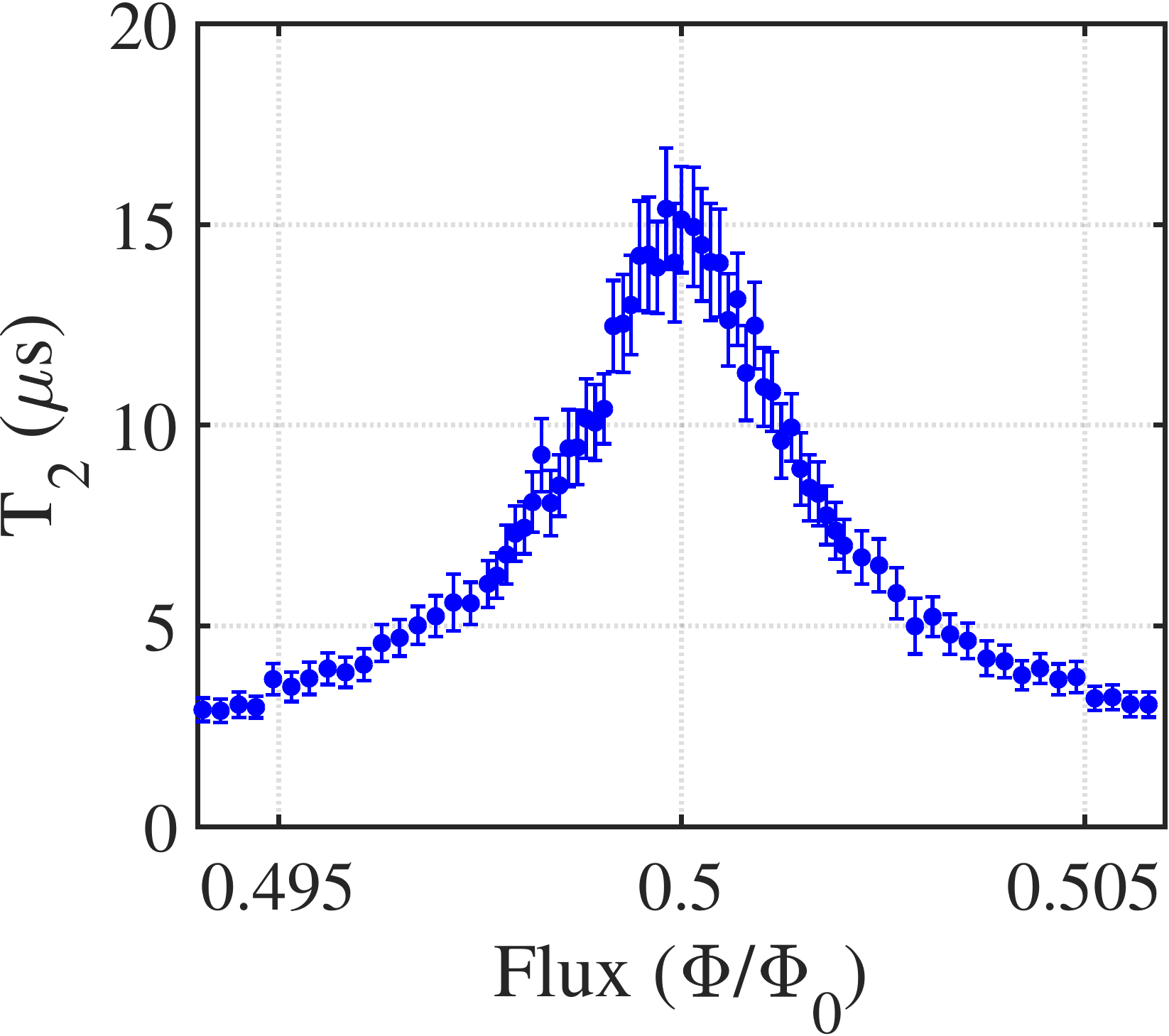}
\end{subfloat}
\caption{\label{fig:CSFQcoherence}CSFQ coherence measured versus flux. (a) $T_1$. (b) $T_2^*$ measured with standard Ramsey sequence. (c) $T_2$ measured with Hahn-echo sequence with a single echo pulse.
}
\end{figure}
In Fig.~\hyperref[fig:CSFQcoherence]{\ref{fig:CSFQcoherence}}, we show the $T_1$, $T_2^*$, and $T_2$ vs. flux for the CSFQ. During the measurements, the $\pi/2$-pulse was recalibrated at each flux point. The $\pi$-pulse was composed of two $\pi/2$ pulses.
\subsection{Single-qubit RB}
\begin{figure}
\makebox[\linewidth]{
\begin{subfloat}{\label{fig:transmon_single_RB}}
\centering
\labellist
\bfseries
\pinlabel (a) at 10 400
\endlabellist
\includegraphics[width=0.3\columnwidth]{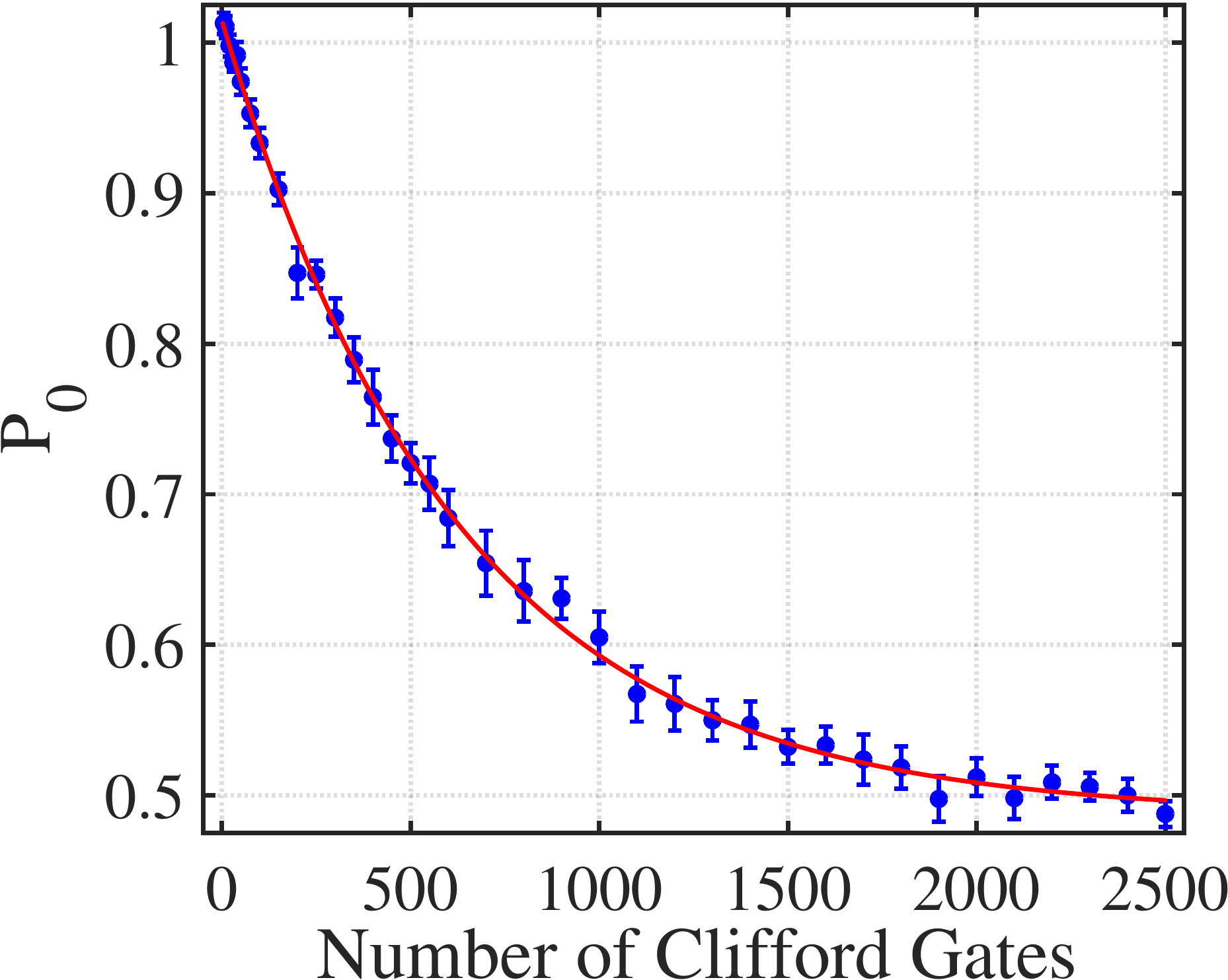}
\end{subfloat}
\begin{subfloat}{\label{fig:CSFQ_single_RB}}
\labellist
\bfseries
\pinlabel (b) at 10 400
\endlabellist
\includegraphics[width=0.3\columnwidth]{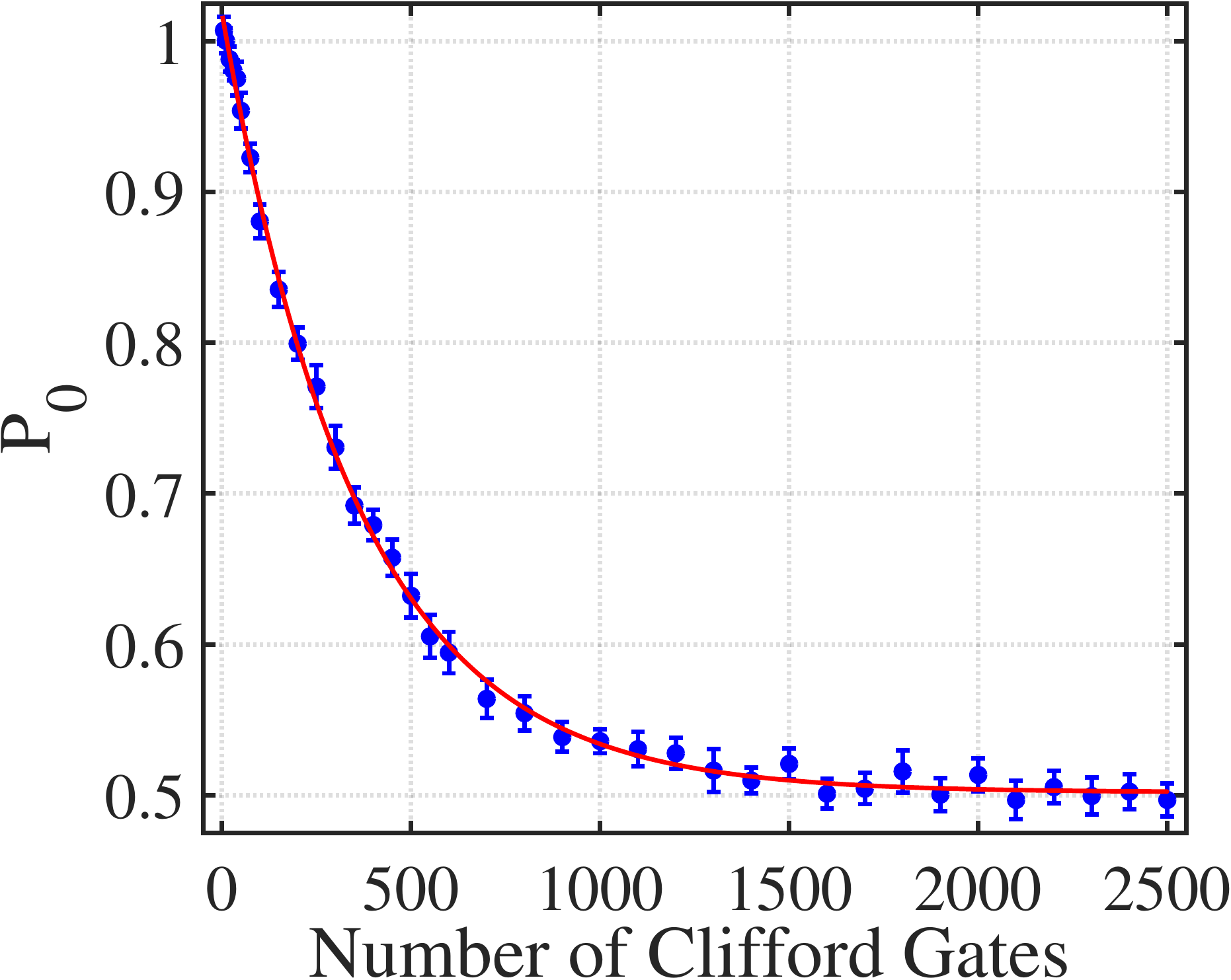}
\end{subfloat}
\begin{subfloat}{\label{fig:two_qubit_RB}}
\labellist
\bfseries
\pinlabel (c) at 10 400
\endlabellist
\includegraphics[width=0.3\columnwidth]{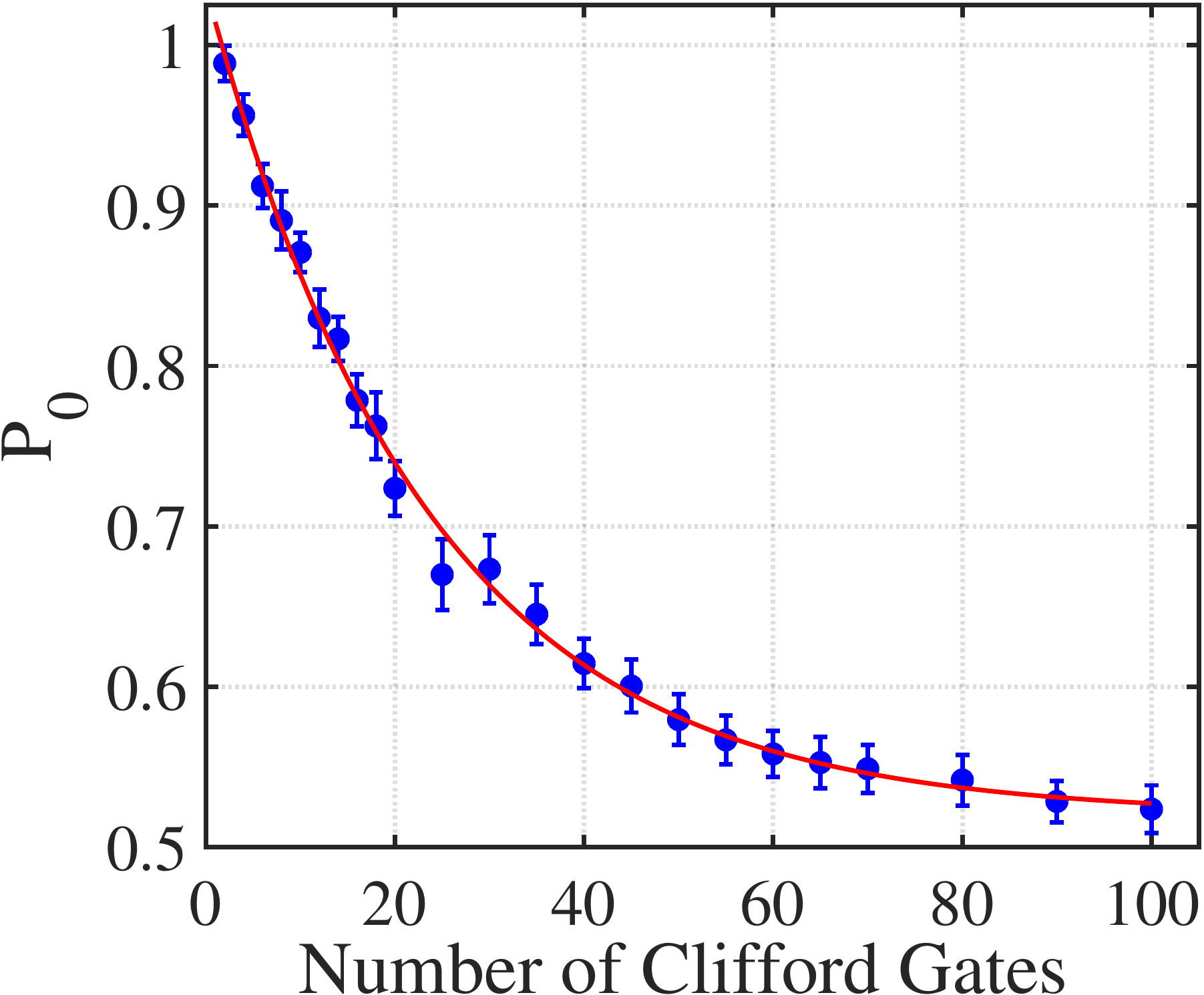} 
\end{subfloat}
}
\caption{\label{fig:RB}Representative RB measurements. (a) single-qubit RB of transmon. (b) single qubit RB of CSFQ at the sweet spot. The single-qubit error per gate is $4.2\times10^{-4}$ ($7.3\times10^{-4}$) for the transmon (CSFQ). (c) Two-qubit RB for 200~ns gate length with the CSFQ at the sweet spot. $P_0$ is the ground state population of the target qubit (transmon). The two-qubit gate error is $1.8\times10^{-2}$. Red solid lines are fit with a fidelity decay function, $A\alpha^m+B$, where $m$ is the number of Clifford gates, $p$ the depolarizing parameter, and A and B the constants that absorb the state preparation and measurement (SPAM) errors. Error bars represent $2\sigma$ confidence intervals.}
\end{figure}
All the gate pulses were generated with single-side-band (SSB) modulation. For single-qubit pulses ($X_{90}$ and $Y_{90}$), we used 20~ns ($=4\sigma$) Gaussian pulses including the derivative removal via adiabatic gate (DRAG) corrections~\cite{Supp:motzoi_simple_2009}, where $\sigma$ is a standard deviation. We use two $X_{90}$ ($Y_{90}$) pulses back-to-back for a $X$ ($Y$) pulse, i.e., $\pi$-pulse. For a pulse calibration of the     $X_{90}$ gate, we first performed a Ramsey measurement to find the qubit transition frequency. We then calibrated the amplitude for the $X_{90}$ pulse via phase estimation~\cite{Supp:kimmel_robust_2015}, followed by the DRAG calibration. The pulse amplitude and DRAG calibration were executed twice to make sure both converged. The $Y_{90}$ pulse was not separately calibrated, but was assumed to have the same amplitude as $X_{90}$.

For single-qubit RB~\cite{Supp:Magesan_RB_PRL2011,Supp:Sheldon_IRB_2016}, we used the pulse primitives $\{I, \pm X_{90}, \pm Y_{90}\}$ to create a group of 24 single-qubit Clifford gates. In RB measurements, we created 30 randomly chosen Clifford sequences for each number of Clifford gates and averaged 2000 times for each pulse sequence to obtain reasonable error bars. By fitting the data to a fidelity decay function of the form $A\alpha^m+B$, we calculated the single-qubit error per gate $\epsilon=1/2\cdot(1-\alpha^{1/N})$, where $N$ is the average number of the pulse primitives in the 24 single-qubit Clifford gates~\cite{Supp:mckay_three-qubit_2019}: $N=2.205$~\cite{Supp:epstein_investigating_2014}.

In Fig.~\hyperref[fig:RB]{\ref{fig:transmon_single_RB}} and~\hyperref[fig:RB]{\ref{fig:CSFQ_single_RB}}, we show two representative RB measurements for the transmon and CSFQ at the sweet spot, and their gate errors. The typical error per gate was lower than $1\times 10^{-3}$ for the transmon and CSFQ over the entire flux range of our experiments. Away from the flux sweet spot, we find that applying the measured $T_1$ and $T_2$ vs. flux significantly overestimates the coherence-limited gate error, consistent with our observations of the flux dependence of the two-qubit gate errors in Fig.~\ref{fig:coh_limit_gate_error}. Again, we attribute this behavior to the nature of RB measurements in the presence of dephasing dominated by $1/f$ noise~\cite{Supp:epstein_investigating_2014}, as discussed earlier.
\subsection{Two-qubit RB}
The pulse primitive for each two-qubit gate is the $ZX_{90}$, which is an echoed CR pulse. CR pulses consist of a flat-top waveform with 20~ns Gaussian rise and fall times ($=2\sigma$). The pulse calibration for the $ZX_{90}$ was performed in two steps: first, we calibrated the phase of the CR pulse so that the rotation axis of the target on the Bloch sphere matches the $x$-axis; next, we calibrated the amplitude of the $ZX_{90}$ gate via phase estimation ~\cite{Supp:kimmel_robust_2015}.

To create the set of two-qubit Clifford gates, we followed Ref.~\onlinecite{Supp:Corcoles_RB_2013}. Each two-qubit Clifford gate was generated from single-qubit primitive gates $\{I, \pm X_{90},\pm Y_{90}\}$ for the transmon and CSFQ, and the two-qubit primitive gate $ZX_{90}$. The ground state probability of the target qubit was measured as a function of the number of randomly chosen two-qubit Clifford gates for each Clifford sequence. We used 20 different random Clifford sequences for each gate length in the RB measurement. For every Clifford sequence, each measurement was averaged 2000 times for obtaining reasonable statistics. As explained in the main text, the gate error $\epsilon$ was calculated by $\epsilon=3/4\cdot(1-\alpha^{1/N})$, where $\alpha$ is the depolarization parameter from the same fidelity decay function as in the single-qubit RB, and $N$ is the average number of two-qubit primitive gates ($ZX_{90})$; $N=1.5$~\cite{Supp:Corcoles_RB_2013,Supp:mckay_three-qubit_2019}. In Fig.~\hyperref[fig:RB]{\ref{fig:two_qubit_RB}} we show a representative two-qubit RB for a 200-ns gate length.
\subsection{Simultaneous RB}
\begin{figure}[h]
\centering
\includegraphics[width=0.45\columnwidth]{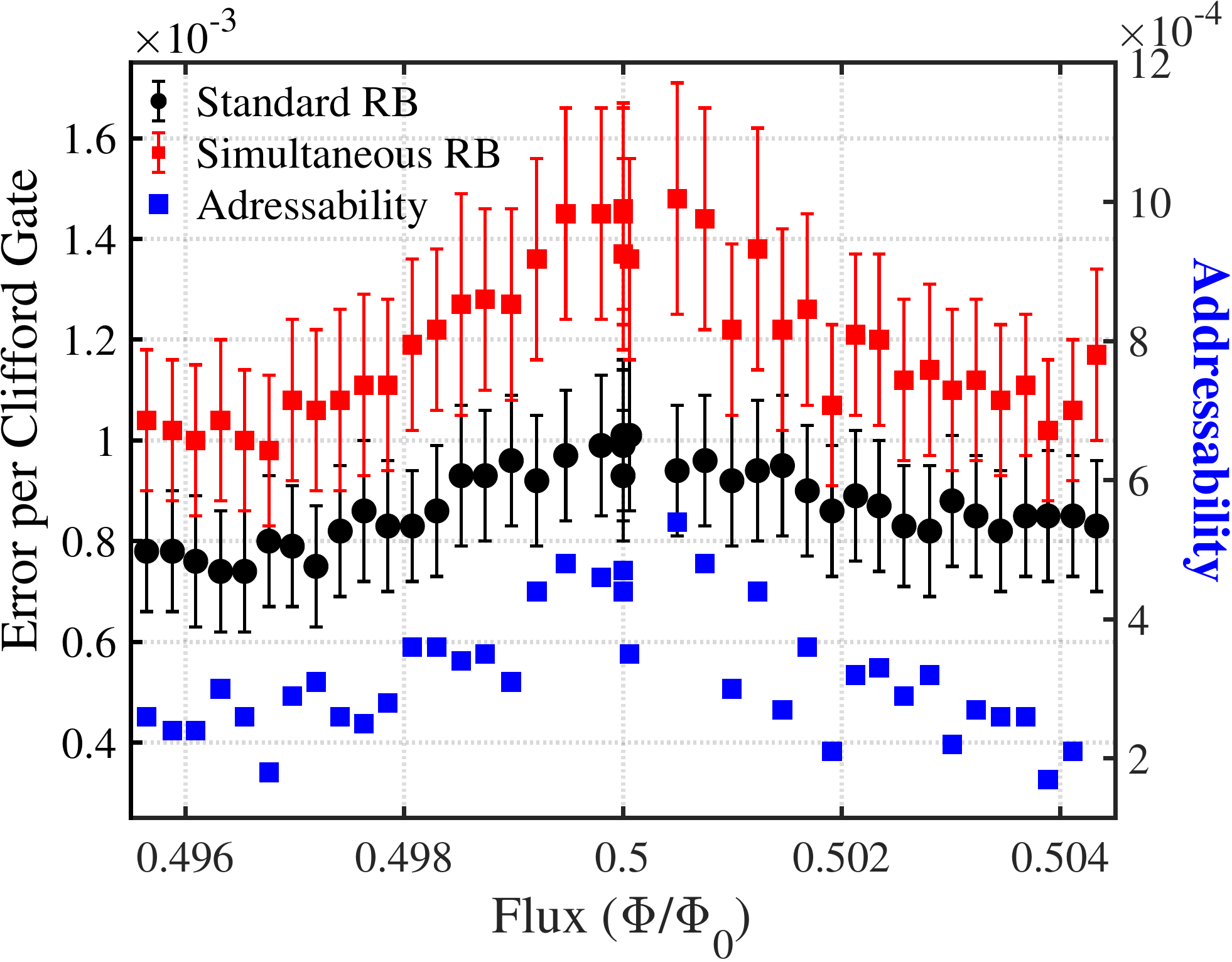}
\caption{\label{transmon_simul_RB}Simultaneous, standard single-qubit RB and addressability for transmon as a function of CSFQ flux bias. An additional error is induced during simultaneous RB. Error bars represent $2\sigma$ confidence intervals.}
\end{figure}
In this section, we show simultaneous single-qubit randomized benchmarking data for the transmon and CSFQ as a function of CSFQ flux bias. The simultaneous RB was performed by applying two different sets of single-qubit RB sequences to the CSFQ and transmon \textit{simultaneously}. Next, we measured each single-qubit RB individually, which, combined with the previous simultaneous RB, allows us to measure the addressability for each qubit. For the transmon, which, again is a fixed-frequency non-tunable qubit, the gate error decreases for CSFQ flux bias points $f\sim0.496$ and 0.504, where $ZZ=0$. This is attributed to the static $ZZ$ interaction, which has a maximum at the CSFQ flux sweet spot. The addressability~\cite{Supp:Gambetta_SimulRB_PRL2012} -- a measure of how much the average error per Clifford gate changes -- is defined by $\delta_{T|C}=r_{T}-r_{T|C}$, where $r_{T}$ is the error per Clifford gate of the transmon without simultaneous RB, and $r_{T|C}$ is the error per Clifford gate of the transmon with simultaneous RB performed on the CSFQ. Clearly, the addressability shows the same dependence on flux as the gate error for the transmon. These results are consistent with the static $ZZ$ measurement and show that the $ZZ$ interaction is a source of error when the two qubits are driven simultaneously, even without performing a two-qubit entangling gate.


\providecommand{\noopsort}[1]{}\providecommand{\singleletter}[1]{#1}%

\end{document}